\documentclass[aps,pra,twocolumn,superscriptaddress,notitlepage,showpacs,showkeys,floatfix]{revtex4-2}

\usepackage{graphicx,amsmath,amsfonts,amssymb,upgreek,txfonts,color,physics}

\usepackage{tabularx}
\usepackage[utf8x]{inputenc} 
\usepackage{color}
\usepackage{import}
\usepackage{enumitem}
\usepackage{lipsum}
\usepackage{mathtools}
\usepackage[dvipsnames]{xcolor}
\usepackage{natbib}
\usepackage[utf8x]{inputenc}
\usepackage{url}
\usepackage[english]{babel}
\usepackage{float}
\usepackage{placeins}
\usepackage{xcolor}
\usepackage{hyperref}

\hypersetup{colorlinks,linkcolor=blue,citecolor=red,urlcolor=blue,breaklinks=true}

\begin{document}

\title{Nonclassical microwave radiation from the parametric dynamical Casimir effect in the reversed-dissipation regime of circuit optomechanics}

\author{H.~Solki} 
\affiliation{Department of Physics, University of Isfahan, Hezar-Jerib, Isfahan 81746-73441, Iran}
\affiliation{Center of Quantum Science and Technology, University of Isfahan, Hezar-Jerib, Isfahan 81746-73441, Iran}

\author{Ali~Motazedifard}
\email{Corresponding author: alimotazedifard@ut.ac.ir $ \& $ motazedifard.ali@gmail.com}
\affiliation{Department of Physics, College of Science, University of Tehran, Tehran 14395‑547, Iran}
\affiliation{Quantum Remote Sensing Lab, Quantum Metrology Group, Iranian Center for Quantum Technologies, Tehran, Tehran 15998-14713, Iran}

\author{M.~H.~Naderi} 
\affiliation{Department of Physics, University of Isfahan, Hezar-Jerib, Isfahan 81746-73441, Iran}
\affiliation{Quantum Optics Group, Department of Physics, University of Isfahan, Hezar-Jerib, Isfahan 81746-73441, Iran}

\author{A.~Youssefi} 
\affiliation{Laboratory of Photonics and Quantum Measurement,
	Swiss Federal Institute of Technology Lausanne, Lausanne, Switzerland}
\affiliation{EDWATEC SA, Lausanne, Switzerland}

\author{R.~Roknizadeh}
\affiliation{Department of Physics, University of Isfahan, Hezar-Jerib, Isfahan 81746-73441, Iran}
\affiliation{Center of Quantum Science and Technology, University of Isfahan, Hezar-Jerib, Isfahan 81746-73441, Iran}
\affiliation{Quantum Optics Group, Department of Physics, University of Isfahan, Hezar-Jerib, Isfahan 81746-73441, Iran}

\date{\today}
\begin{abstract}
We propose an experimentally feasible optomechanical system (OMS) that is dispersively driven and operates in the reversed dissipation regime (RDR), where the mechanical damping rate far exceeds the cavity decay rate. We demonstrate that coherent, fast-time modulation of the driving laser frequency-on time scales longer than the mechanical decoherence time-allows for adiabatic elimination of the mechanical mode, resulting in strong parametric amplification of quantum vacuum fluctuations of the intracavity field. This mechanism, known as the parametric dynamical Casimir effect (parametric-DCE), leads to the generation of Casimir photons. In the dispersive RDR, we find that the total system Hamiltonian—including the DCE term—is intrinsically modified by a generalized optomechanical Kerr-type nonlinearity. 
This nonlinearity not only saturates the mean number of radiated Casimir photons on short time scales, even without dissipation, but also induces oscillatory behavior in their dynamics and quantum characteristics. 
Remarkably, the presence of the Kerr nonlinearity causes the generated DCE photons to exhibit nonclassical features, including simultaneous sub-Poissonian statistics and negative Wigner function, as well as quadrature squeezing which can be controlled by adjusting the system parameters. Surprisingly, the controllable simultaneous nonclassical dynamics in the same physical parameter regime which is induced by the optomechanical Kerr nonlinearity to the parametric DCE cannot occurs in the standard DCE or Kerr-type systems.
The proposed nonclassical microwave radiation source possesses the potential to be applied in quantum information processing, quantum computing as well as microwave quantum sensing.
\end{abstract}

\hbadness=10000
\maketitle

\section{Introduction}

It is well-known that quantum parametric processes provide essential mechanisms for generating and manipulating quantum states in diverse aspects of quantum science and technologies. The most remarkable property of these processes concerns the dynamical amplification of quantum vacuum fluctuations that is responsible for the creation of pairs of real particles as a consequence of strong non-adiabatic change of a system parameter or boundary condition \cite{Moore1970,Davies1976}.
This effect, which is generally referred to as the dynamical Casimir effect (DCE) \cite{Moore1970,Yablonovitch1989,Schwinger1992,Nation2012,Dodonov2020},
describes a process in which a cavity with periodically oscillating mirrors produces pairs of photons out of the electromagnetic vacuum fluctuations. Along with the theoretical studies on the issue of the possibility of particle creation via  the DCE in a large variety of systems, ranging from cosmology \cite{Brevik2000,Durrer2007,Lock2017} to non-stationary cavity QED \cite{Dalvit1999,Crocce2002,Dodonov2003,Dodonov2011,Mendoza2019,Lo2020},
various theoretical schemes for practical applications of the DCE have been suggested, including generation of photons with nonclassical properties \cite{Dodonov1990,Dodonov1999,Johansson2013}, generation of squeezed \cite{Aggarwal2015}, and entangled atomic states \cite{Lange2018}, generation of multipartite entanglement in networks of quantum cavities and superconducting devices \cite{Felicetti2014}, generation of entanglement between two moving qubits \cite{Agusti2019},
generation of EPR quantum steering and Gaussian interferometric power \cite{Sabin2015}, and 
quantum synchronization of two qubits \cite{Mitarai2024}.
In addition, in the context of quantum thermodynamics, it has been shown that the DCE can be used for realizing the so-called anti-DCE \cite{Sousa2015}, i.e., coherent annihilation of photons instead of creation, as a resource for work extraction from the atom-field system \cite{Dodonov2017}, for implementation of a quantum Otto heat engine \cite{Ferreri2023}, and for cooling down the cavity wall in the presence of a non-vanishing temperature gradient between the wall and the cavity \cite{Ferreri2024}.

From an experimental point of view, for a measurable flux of real photons to be generated, the moving boundaries should oscillate at very high frequencies (more than 10GHz) that are not yet experimentally achievable. Consequently, alternative schemes based on imitation of boundary motion have been proposed, with examples including periodic modulation of the optical properties of the boundary \cite{Lombardi, Dodonov2006, Dodonov2006-2} or of the optical path length of a cavity \cite{Dezael,Faccio,Motazedifard DCE}, amplification of the DCE in a cavity within the Rabi model in the ultrastrong coupling regime via quantum optimal control strategies \cite{Hoeb}, and time modulation of the Kerr or higher-order nonlinearities in a cavity \cite{Dodonov2022}.
Some other experimental schemes aiming at the observation of the DCE can be found in Refs.~\cite{Lombardi,Agnesi,Kawakubo}. 
Notably, it has been reported about a successful implementation of DCE in superconducting circuits through two independent experiments by fast-modulating either the electrical boundary condition of a transmission line \cite{Pourkabirian} or the effective speed of light in a Josephson metamaterial \cite{Lahteenmaki}.
Moreover, recently an analogue DCE has been experimentally realized in the near infrared regime using a dispersion-oscillating photonic crystal fiber \cite{vezzoli2019}. Results of this experiment implied that the generated DCE photons exhibit nonclassical photon anti-bunching statistics.

Besides the photonic DCE, analog models for the generation of other particles than photons via the DCE have also been investigated such as the DCE of phonons in a time-modulated atomic Bose-Einstein condensate (BEC) \cite{Recati,Jaskula}, emission of Bogoliubov excitations in an exciton-polariton condensate by an ultrashort laser pulse \cite{Koghee}, phononic DCE in a time-modulated quantum fluid of light \cite{Busch}, DCE of magnon excitation in a spinor BEC driven by a time-dependent magnetic field \cite{Saito}, and DCE of phonons in a gas of laser-cooled atoms with time-dependent effective charge \cite{Dodonov2014}.

On the other hand, during the recent decades optomechanical systems (OMSs) where electromagnetic radiation pressure is linearly or quadratically coupled to a mechanical oscillator (MO) \cite{Aspelmeyer2014} have been widely employed in a large variety of applications, for example, displacement and force sensing \cite{Kippenberg2007,Tsang2010,Tsang2012,Bariani2015,Wimmer2014,Motazedifard2016,Allahverdi2022,Motazedifard2021,Motazedifard2019,Ebrahimi2021,Bemani2022}, optomechanical cooling \cite{Connell2010,Teufel2011,Chan2011}, quantum correlations such as entanglement \cite{Palomaki2013,Paternostro2007,Genes2009,Dalafi2018,Bemani2019} and synchronization \cite{Mari2013,Ying2014,Bemani2017}, generation of nonclassical states of the mechanical and optical modes \cite{Nunnenkamp2011,Jahne2009,Latmiral2018,Groblacher2018,Brunelli2018,Solki2023}, quadrature squeezing/amplification \cite{Kippenberg2008,Massel2017,Ockeloen2018,Motazedifard2019-2}, optomechanically induced transparency \cite{Xiong2018,Mikaeili2022,Agarwal2010,Weis2010,Motazedifard2021-2,Safavi2011,Karuza2013,Kronwald2013,Motazedifard2023,Yang2019}, and quantum illumination radar \cite{Barzanjeh2015,Barzanjeh2020,Allahverdi2024}.

In recent years, a variety of theoretical studies have identified the OMSs as promising platforms for realizing the DCE. To mention some of them, one could highlight the DCE of phonons in a non-stationary quantum-well assisted optomechanical cavity \cite{Mahajan2015}; the DCE of phononic excitation in the so-called membrane-in-the-middle optomechanical cavity \cite{Motazedifard2017}; the DCE of photons and mechanical-/Bogoliubov-type phonons in a modulated hybrid optomechanical cavity containing an atomic BEC \cite{Motazedifard2018}; the DCE of photons in an optomechanical cavity under incoherent mechanical excitation \cite{Settineri2019}; the DCE of photons in a squeezed-cavity-assisted OMS \cite{Qin2019}; the DCE of photons in an optomechanical cavity interacting with a one-dimensional photonic crystal \cite{Tanaka2020}; the DCE of phonons in the ultrastrong-coupling regime of optomechanics \cite{Minganti2024}; and the DCE of photons in non-perturbative coupling regime of cavity optomechanics \cite{Macri2018}. Remarkably, in Ref.~\cite{Macri2018} the authors have found that the standard resonance condition for the generation of DCE photons which requires the mechanical frequency to be, at least, twice the first cavity mode frequency does not need to be satisfied, provided that the coupling between the cavity field and the moving mirror is non-vanishing with respect to the optical and mechanical resonance frequencies. In addition, they have shown that the non-perturbative regime of DCE can give rise to the steady-state entanglement between the moving mirror and the cavity field.

One of the most important characteristics of cavity OMSs is a type of intrinsic nonlinearity resulting from the radiation pressure coupling. This nonlinearity stems from the fact that in a typical cavity OMS the position of the MO modulates the resonance frequency of the cavity mode. In other words, the optical length of the cavity depends on the intensity of the intracavity field and, consequently, the optomechanical cavity can behave effectively as a nonlinear Kerr medium \cite{Gong2009,Aldana2013}.
This inherent nonlinearity enables pondermotive squeezing of the cavity field \cite{Fabre1994,Mancini1994}, generation of mechanical cat states \cite{Bose1997,Bose1999}, intracavity cat-state generation \cite{Leghtas2015} photon blockade \cite{Rabl2011,Ling2023}, optical bistability \cite{Dorsel1983,Gozzini1985}, phonon–photon entanglement in the bistable regime \cite{Ghobadi2011}, and cooling an optomechanical resonator into a cat state of motion \cite{Hauer2023}.

The above-mentioned cavity optomechanical phenomena, such as mechanical ground-sate cooling, entanglement, quadrature squeezing/amplification, and optomechanically induced transparency emerge in the so-called normal dissipation regime, in which the decay rate of the MO is typically significantly smaller compared to that of the cavity mode. However, recent investigations in the field of dissipation engineering have introduced a novel regime of dissipation in optomechanics, the so-called reversed-dissipation regime (RDR), where the roles of the cavity mode and the MO are interchanged \cite{Nunnenkamp2014}, i.e., the dissipation rate of the MO prominently exceeds the cavity linewidth. In this regime, the MO plays the advantageous role of an additional cold dissipative reservoir for the cavity mode \cite{Wang2013,Metelmann2014}. Recently significant development  in experimental implementations has made possible the realization of the RDR in various experimental setups, including superconducting circuits \cite{Sliwa2015,Lecocq2017}, microwave-cavity optomechanical systems \cite{Toth2017}, and mechanical resonators embedded by erbium ions \cite{Ohta2021}. In addition, a variety of applications have been proposed and investigated for this regime. Some examples include quantum-limited amplification and self-oscillation of photons \cite{Nunnenkamp2014,Toth2017}, entangled-photon generation \cite{Wang2013}, nonreciprocal optomechanically induced transparency and enhancing  the ground-state cooling of a MO \cite{Zhang2024}, and broadband nonreciprocal and chiral photon transmission \cite{Chen2023}.

Inspired by the above-mentioned investigations, in this paper, we propose a protocol to study the parametric DCE in an experimentally realized microwave cavity OMS \cite{Toth2017} which is dispersively driven and operates in the RDR. In such case, the degrees of freedom of the MO can be adiabatically eliminated from the system dynamics. We show that the coherent time modulation of the driving laser frequency leads to the parametric amplification of the quantum vacuum fluctuations of the intracavity field mode, resulting in the creation of microwave Casimir photons over time scales longer than the mechanical decoherence time. In addition, the adiabatic elimination of the mechanical mode will induce an effective nonlinear Kerr-type photon–photon interaction, which provides the feasibility of controllable manipulation of the quantum properties of the generated DCE photons. We analyze both analytically and numerically the effects of the induced Kerr nonlinearity as well as the modulation amplitude of the driving laser frequency on the dynamical behaviors of the mean number of the generated microwave Casimir photons, photon counting statistics, and photon quadrature squeezing. Within the short-time approximation, i.e., over time scales much shorter than the cavity decay time, we find a closed analytical expression for the system unitary time-evolution operator by using the Wei-Norman theorem \cite{Norman1964}, which is of the form of a generalized squeezing operator affected by the induced Kerr nonlinearity. The presence of the Kerr nonlinearity causes that the time evolution of the mean number of Casimir photons exhibits an oscillatory behavior whose amplitude decreases with increasing time. Regarding the nonclassical features of the generated DCE photons, we find that they can exhibit quadrature squeezing and sub-Poissonian statistics in the course of time evolution. 

Beyond these general dynamical features, our work identifies several aspects that distinguish this scheme from previous DCE studies. We show that the generated Casimir photons can simultaneously exhibit sub-Poissonian statistics and Wigner-function negativity, a combination of nonclassical signatures not previously reported in DCE settings. Moreover, the proposed DCE+Kerr mechanism relies on experimentally accessible circuit-optomechanical parameters in the RDR, ensuring practical implementability. Finally, the interplay between the parametric modulation and the \textit{optomechanically}-induced Kerr nonlinearity gives rise to additional dynamical signatures such as oscillatory and saturation behaviour in photon generation and periodic redistribution of squeezing between quadratures demonstrating quantum features beyond those achievable in standard DCE or fixed-Kerr models.

The paper is structured as follows. First, we introduce the physical system under consideration in Sec.~\ref{Section2}, and derive an effective system Hamiltonian in the RDR of cavity optomechanics when the cavity is dispersively driven by an external modulated field. The effects of the induced Kerr nonlinearity on the dynamics of the mean number, the quantum statistics, and the quadrature squeezing of the generated Casimir photons are, respectively, discussed in Secs.~\ref{Section3},~\ref{Section4} and \ref{Section5}. 
In Sec.~\ref{Section6} we explore the signature of the nonclassical characteristic of the generated Casimir photons by calculating the corresponding Wigner function. In particular, we examine the impact of system dissipation on the temporal behavior of the Wigner negativity. 
A detailed discussion of the experimental requirements for implementation of the model is undertaken in Sec.~\ref{Section7}.  We summarize the conclusions of our work in Sec.~\ref{Section8}. 
Some detailed mathematical calculations can be found in the appendices.


\section{Physical Model and The Effective System Hamiltonian}\label{Section2}

\begin{figure}
	\includegraphics[width=8.6cm]{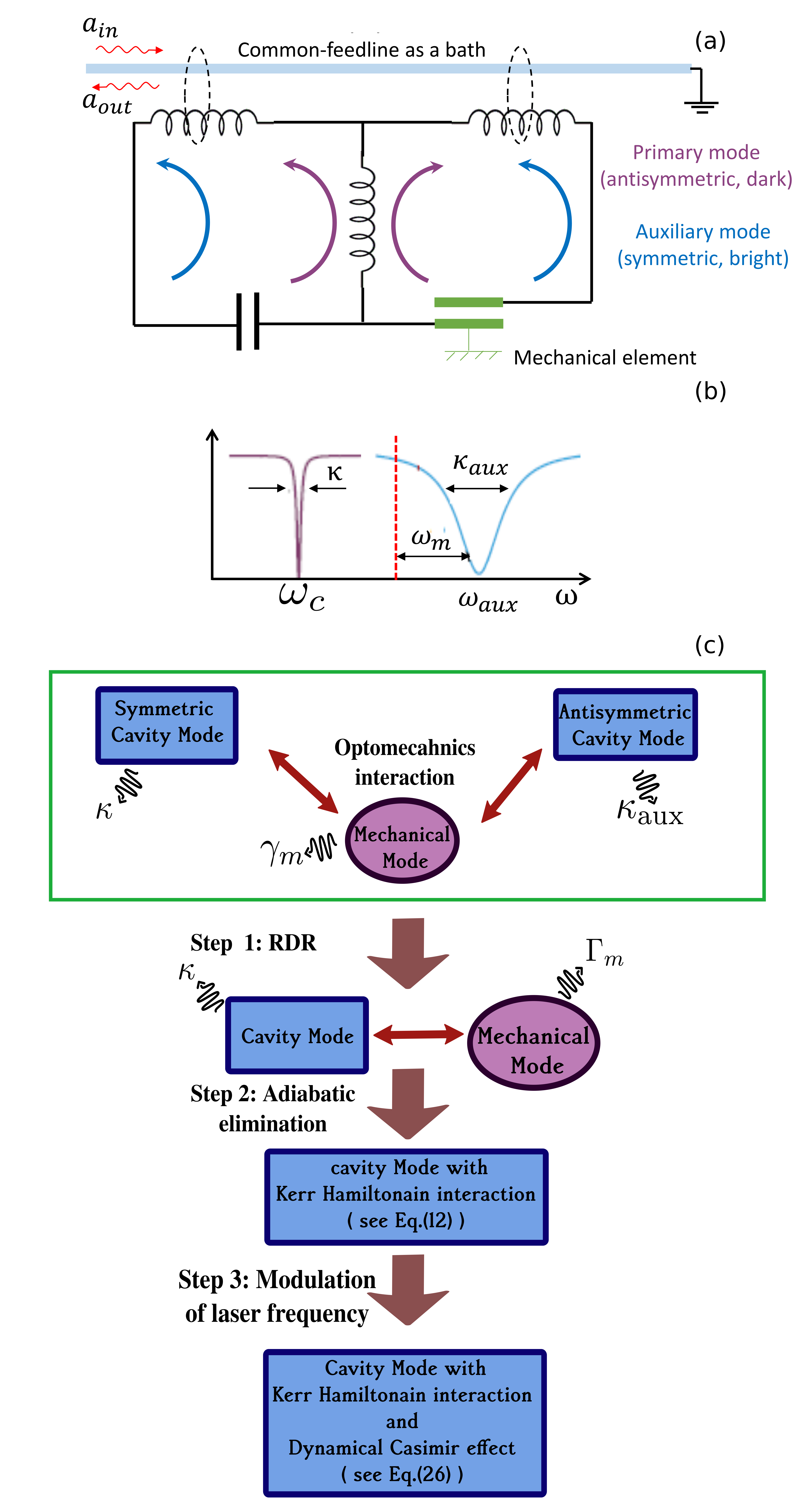}
	\caption{(Color online) 
		(a) Schematic diagram of the considered microwave optomechanical circuit \cite{Toth2017} composed of two inductively coupled microwave LC resonators which are inductively coupled to a microwave feedline as a common bath. The two hybridized dark and bright modes of the circuit that are, respectively, used as the primary $\hat{a}$ and auxiliary $\hat{a}_{\rm aux}$ interact with the vibration of the suspended top electrode of a shared capacitor, acting as a MO with resonance frequency $\omega_m$. 
		The auxiliary (primary) mode is driven by a classical laser field of frequency $\omega_L^{\rm aux}(\omega_L)$ and amplitude $E_L^{\rm aux}(E_L)$. (b) To prepare a cold, dissipative mechanical reservoir for microwave photons the dissipation rate $\gamma_m$ of the MO should be increased to match the much larger dissipation rate $\kappa$ of the primary mode having resonance frequency $\omega_c$. 
		For this purpose, the auxiliary mode with resonance frequency $\omega_{\rm aux}$ and dissipation rate $\kappa_{\rm aux}$ is used to damp out the MO via optomechanical sideband cooling and, hence, prepare it as a strongly dissipative, cold reservoir for the primary mode. 
		(c) A schematic overview of the process to derive Hamiltonian~\eqref{H_DCE2}. }
	\label{fig1}
\end{figure}

As schematically illustrated in Fig.~\ref{fig1}(a), the system we have chosen to consider is a microwave-cavity optomechanical circuit, which has been previously presented in Ref.~\cite{Toth2017} aiming at realizing a cold, dissipative mechanical reservoir for microwave photons. It is composed of two inductively coupled microwave LC resonators, both coupled inductively to a common microwave feedline, serving as a common bath, with external coupling strengths $\kappa_1^{\rm ex}$ and $\kappa_2^{\rm ex}$. One of the LC resonators contains a mechanically compliant capacitor that can be modeled as a single-mode MO. By making use of an electromagnetic mode, referred to as the auxiliary mode, the MO is damped out via optomechanical sideband cooling~\cite{Schliesser2008} in order to prepare it as a strongly dissipative, cold reservoir for another electromagnetic mode which referred to as the primary mode.

In order to accomplish the RDR in this setup, the auxiliary mode needs to be cooled down with a rate much faster than the electromagnetic decay rate of the primary mode, requiring the microwave cavities to have extremely different decay rates. One successful approach to overcome this challenge is to engineer hybridized modes with intrinsically different decay rates arising from interference in the output channel [for details, see the Supplementary Information in Ref.~\cite{Toth2017}].
The resulting interaction Hamiltonian is given by \cite{Toth2017} ($\hbar=1$)
\begin{eqnarray} \label{H_RDR1}
&& \hat H_{\rm int}=  J (\hat a_1^\dag \hat a_2  + \hat a_2^\dag \hat a_1) - g_0' \hat a_1^\dag \hat a_1 (\hat b+ \hat b^\dag) ,
\end{eqnarray}
where $\hat{b}$ is the annihilation operator for the mechanical mode with frequency $\omega_m$ and decay rate $\gamma_m$, $\hat{a}_1$ and $\hat{a}_2$ are the annihilation operators for the bare modes with frequency $\omega_0$, $J$ denotes the intermode coupling strength, and $g_0'$ stands for the vacuum electromechanical coupling strength to the first bare mode. 
By introducing the symmetric and antisymmetric superpositions of the bare modes $ \hat a_{s,a} = (\hat a_1 \pm \hat a_2)/\sqrt{2}$ \cite{Toth2017}  the intermode coupling term can be diagonalized, and thus the interaction Hamiltonian is expressed as
\begin{eqnarray}\label{H_RDR2} 
\hat H_{\rm int}=   J (\hat a_s^\dag \hat a_s   - \hat a_a^\dag \hat a_a)  - \frac{g_0'}{2} (\hat a_s^\dag \hat a_s +  \hat a_a^\dag \hat a_a) (\hat b  +  \hat b^\dag).
\end{eqnarray}

The above Hamiltonian is written in the limit of large intermode coupling compared to the mechanical frequency, $ J \gg \omega_m $, which allows to neglect the cross terms $ \hat a_s^\dag \hat a_a(\hat b+ \hat b^\dag)$ as well as the other non-resonance terms.
Although the bare modes are degenerate, the primary and auxiliary eigenmodes with respective energies of $\omega_c=\omega_0-J$ and $\omega_{\rm aux}=\omega_0+J$ have an energy difference of $2J$ for $J\neq 0$. As is evident from the Hamiltonian~\eqref{H_RDR2}, the intermode coupling $(J\neq 0)$ results in that the eigenmodes are both coupled to the MO with coupling strength $g_0'/2$.

The hybridized modes $\hat{a}_{s,a}$ constitute a strongly coupled (bright) and a weakly coupled (dark) mode with the respective decay rates $\kappa_s^{\rm ex}=\kappa_1^{\rm ex}+\kappa_2^{\rm ex}$ and $\kappa_a^{\rm ex}=\abs{\kappa_1^{\rm ex}-\kappa_2^{\rm ex}}$, which can be explained as the result of the interference of the bare-mode external coupling rates $\kappa_1^{\rm ex}$ and $\kappa_2^{\rm ex}$  with the output channel [see Ref.~\cite{Toth2017} and its Supplementary Information for details]. 
In the case of nearly the same coupling rates for the bare modes ($\kappa_1^{\rm ex}\approx\kappa_2^{\rm ex}$) we have $\kappa_s^{\rm ex}\gg\kappa_a^{\rm ex}$ so that the antisymmetric (dark) mode is approximately decoupled from the common reservoir. 
Following Ref.~\cite{Toth2017}, in subsequent discussions we refer the dark and the bright modes as the primary and the auxiliary modes, respectively, with respective resonance frequencies $\omega_c$ and $\omega_{\rm aux}$ and decay rates $\kappa$ and $\kappa_{\rm aux}$ [see Fig.~\ref{fig1}(b)].

The total Hamiltonian of the system can be written as
\begin{equation}\label{H_total}
\hat{H}=\hat{H}_{m} + \hat{H}_{\rm pr} + \hat{H}_{\rm aux},
\end{equation}
with
\begin{subequations}
\begin{equation}
\hat{H}_{m}=\omega_m \hat{b}^\dagger\hat{b},
\end{equation}
\begin{equation}
\hat{H}_{\rm pr}= \omega_c \hat{a}^\dagger\hat{a} -g_0\hat{a}^\dagger\hat{a}(\hat{b}^\dagger+\hat{b})+iE_L(\hat{a}^\dagger e^{-i\omega_L t}-\hat{a}e^{i\omega_L t}),
\end{equation}
\begin{equation}
\hat{H}_{\rm aux}= \omega_{\rm aux} \hat{a}_s^\dagger\hat{a}_s -g_0\hat{a}_s^\dagger\hat{a}_s(\hat{b}^\dagger+\hat{b})+iE_L(\hat{a}_{\rm aux}^\dagger e^{-i\omega_L^{\rm aux} t}-\hat{a}_{\rm aux}e^{i\omega_L^{\rm aux} t}).
\end{equation}
\end{subequations} 
The Hamiltonian $\hat{H}_m$ is the free Hamiltonian of the MO and the three terms in $\hat{H}_{\rm pr}(\hat{H}_{\rm aux})$ denote, respectively, the free Hamiltonian of the primary (auxiliary) mode, the optomechanical coupling of the primary (auxiliary) mode to the MO with coupling strength $g_0=g_0'/2$, and driving the primary (auxiliary) mode with the classical laser field of frequency $\omega_L(\omega_L^{\rm aux})$ and amplitude $E_L(E_L^{\rm aux})$.  
Here, an important point to remark is that the Hamiltonian of Eq.~\eqref{H_total} consists of the extra term $g_0(\hat{a}^{\dagger 2}+\hat{a}^2)(\hat{b}+\hat{b}^\dagger)$, which is induced by the mechanical motion and responsible for the traditional DCE~\cite{Macri2018}. However, in the most common experimental situation of cavity optomechanics, where the mechanical frequency is much smaller than the cavity frequency, this term can be neglected with good approximation. In fact, the experimental generation of a significant number of real photons from vacuum, sufficient to allow detection, requires the mechanical frequency to be at least twice that of the cavity, which still remains a serious problem. Recently a high mechanical frequency as large as $\omega_m/2\pi\sim6$GHz has been reported \cite{Connell2010,Rouxinol2016} using microwave resonators and ultra-high-frequency mechanical micro- or nano-resonators. 
However, for the generation of Casimir radiation at the frequency of about $5$GHz a still higher mechanical frequency, $\omega_m/2\pi\sim10$GHz is needed. As mentioned before to bypass this experimental difficulty, alternative schemes based on the simulation of the high-frequency mechanical motion can be exploited to produce the parametric DCE. 
This is the case addressed in the present paper, where the parametric DCE arises from the modulation of the driving laser frequency.

As explained before, in order to prepare the MO in the RDR of optomechanics it should be cooled down by the auxiliary cavity mode possessing a large damping rate $\kappa_{\rm aux}\gg(\gamma_m,\kappa)$. After cooling the MO, the auxiliary cavity mode can be adiabatically eliminated. Therefore, on time scales greater than $\kappa_{\rm aux}^{-1}$ and in the frame rotating with the driving laser frequency $\omega_L$ the total Hamiltonian of the system can be expressed as (see Appendix~\ref{Appendix_A} for details of derivation) 
\begin{equation}\label{H2}
\hat H= -\Delta_c \hat a^\dag \hat a  + \omega_m \hat b^\dag \hat b - g_0 \hat a^\dag \hat a (\hat b +\hat b^\dag) + i E_L (\hat a^\dag -\hat a),
\end{equation}
where $\Delta_c=\omega_L - \omega_c$ is the detuning between the primary mode and its driving field.

The dynamics of the system governed by the effective Hamiltonian~\eqref{H2} is fully characterized by the following set of Langevin equations,
\begin{subequations} 
	\begin{eqnarray}
	\label{langevin1}
	&& \dot {\hat a} =  i \Delta_c \hat a + i{g_0}\hat a (\hat b + {{\hat b}^\dag }) + {E_L} - \frac{\kappa }{2}\hat a + \sqrt \kappa  {{\hat a}_{\rm in}}, \\
	\label{langevin1-2}
	&& \dot {\hat b }=  - i{\omega _m}\hat b + i{g_0}{{\hat a}^\dag }\hat a  - \frac{{{\Gamma_m}}}{2}\hat b + \sqrt {{\Gamma _m}}  \hat{\tilde b}_{\rm in},
	\end{eqnarray}
\end{subequations} 
where is a generalized mechanical noise operator whose explicit expression $\hat{\tilde{b}}_{\rm in}$ is given by Eq.~\eqref{b_in}, and $\hat{a}_{\rm in}$ stands for the input vacuum noise for the primary mode characterized by the non-vanishing correlation function $\langle {{{\hat a}_{\rm in}}(t)\hat a_{\rm in}^\dag (t')}\rangle = \delta (t - t')$. 

Since the MO operates in the RDR of cavity optomechanics where $\Gamma_m \gg \kappa$, the mechanical mode can be adiabatically eliminated on time scales much longer than $\Gamma_m^{-1}$. To this end, one can formally integrate Eq.~\eqref{langevin1-2} to obtain
\begin{eqnarray} \label{b2}
&& \hat b(t) \approx g_0 \frac{\omega_m + i \Gamma_m/2}{\omega_m^2 + \Gamma_m^2/4} \hat a^\dag \hat a + \sqrt{\Gamma_m } \frac{\Gamma_m/2 - i \omega_m}{\omega_m^2 + \Gamma_m^2/4} \hat {\tilde b}_{\rm in}(t).
\end{eqnarray}
In the limit of high mechanical quality factor, $ \omega_m \gg \Gamma_m $, and in the interaction picture Eq.~\eqref{b2} takes the form
\begin{eqnarray} 
\hat b(t) \approx \frac{g_0}{\omega_m} \hat a^\dag \hat a -i \frac{\sqrt{\Gamma_m }}{\omega_m}  \hat{\tilde b}_{\rm in}(t). \label{bfinal} 
\end{eqnarray}
Substituting this equation into Eq.~\eqref{langevin1} leads to the following equation for the primary cavity mode
\begin{eqnarray} 
\dot {\hat a} =  i \Delta_c \hat a + 2i \frac{g_0^2}{\omega_m} \hat a \hat a^\dag \hat a + {E_L} - \frac{\kappa }{2}\hat a + \sqrt \kappa \hat F_{\rm in}(t),  \label{langevin2}
\end{eqnarray}
where the generalized nonlinear Markovian noise $ \hat F_{\rm in}(t) $ is given by 
\begin{equation}\label{F_noise}
\hat F_{\rm in}(t) = \hat  a_{\rm in} - \sqrt{\frac{\Gamma_m}{\kappa}} \frac{g_0 \hat a}{\omega_m} (\hat {\tilde b}_{\rm in}^\dag - \hat {\tilde b}_{\rm in}),
\end{equation}
obeying the correlation function
\begin{gather}
\langle{ \hat F_{\rm in}(t)\hat F^\dagger_{\rm in}(t')}\rangle=\delta(t-t')\Big[
1 + \frac{g_0^2}{\kappa \omega_m^2} n_c \Big(\gamma_m(2\bar{n}_{\rm m}+1) 
\nonumber
\\
+ \frac{4}{\kappa_{\rm aux}} 
+ \abs{\frac{\kappa_{\rm aux}/2-2i\Delta_{\rm aux}}{\kappa^2_{\rm aux}/4+4\Delta_{\rm aux}^2}}^2 
+ \frac{2\sqrt{\kappa_{\rm aux}}}{\kappa^2_{\rm aux}/4+4\Delta_{\rm aux}^2}
\Big)
\Big],
\end{gather}
with $ n_c= \langle \hat a^\dagger \hat a \rangle(t) $. Equation \eqref{F_noise} shows that the input noise $\hat{a}_{\rm in}$ acts as an additive noise on the primary cavity mode. 
In contrast, the noise term induced by the adiabatic elimination of the mechanical mode (second term in Eq.~\eqref{F_noise}) is a multiplicative noise. For time scales much shorter than the characteristic time of the primary cavity mode, $t\ll \kappa^{-1}$, one can neglect the impact of damping and noise on the cavity mode $\hat{a}$. This is the case that we consider in what follows. 

From Eq.~\eqref{langevin2}, one can find the effective nonlinear Hamiltonian of the system as follows
\begin{equation}\label{H_eff}
\hat H_{\rm eff} =   -\bar{\Delta}_c \hat a^\dag \hat a  -  g_K \hat a^{\dag 2} \hat a^2  +  i E_L (\hat a^\dag  -\hat a),
\end{equation}
where the second term represents the optomechanically-induced Kerr nonlinearity for the primary cavity mode with $g_K= g_0^2/\omega_m$ playing the role of the third-order susceptibility ($ \chi_{\rm eff,kerr}^{(3)} $) as in usual Kerr media.
Furthermore, $ \bar \Delta_c = \Delta_c + 2g_K$ stands for the shifted cavity detuning which can be considered as an effective frequency for the primary mode in the RDR.

It is important to note that the Kerr-type term in Eq.\eqref{H_eff} is fundamentally different from the standard Kerr interaction in conventional $\chi^{(3)}$ media. In ordinary Kerr materials, the nonlinearity is fixed by intrinsic material properties and is therefore not tunable. In our case, the Kerr interaction arises \emph{effectively} from the adiabatic elimination of the mechanical mode in the RDR of optomechanics, making its strength $g_{K}$ fully controllable through optomechanical coupling, which constitutes an advantage not available in standard Kerr platforms.


\subsection{Simulation of the parametric DCE Hamiltonian}
Having obtained the effective system Hamiltonian~\eqref{H_eff} in the RDR of cavity optomechanics, we are now in a position to show the possibility of simulation of the parametric DCE using frequency modulation of the laser driving the primary mode. To do this, first, it is convenient to rewrite Hamiltonian~\eqref{H_eff} as
\begin{equation}\label{H_eff_r}
\hat H_{\rm eff} =   -\bar{\Delta}_c 
\big[
\hat a^\dag \hat a  +  C_K \hat a^{\dag 2} \hat a^2  -  i C_E (\hat a^\dag  -\hat a)
\big],
\end{equation}
where
\begin{equation}\label{C}
C_E=\frac{E_L}{\bar\Delta_c} ,\quad C_K=\frac{g_K}{\bar\Delta_c}.
\end{equation}
In the dispersive regime, i.e., when the detuning $\Delta_c$ is much larger than any frequencies in the system, the driving term in the effective Hamiltonian~\eqref{H_eff_r} can be approximately removed by applying the unitary transformation
\begin{align}\label{H_eff_D}
\hat{\tilde{H}}_{\rm eff} = \hat{D}\hat H_{\rm eff}\hat{D}^\dagger 
\approx
-\bar\Delta_c[\hat a^\dag \hat a  +  C_K \hat a^{\dag 2} \hat a^2],
\end{align}
where
\begin{equation}\label{D}
\hat{D}=\exp(-iC_E(\hat{a}^\dagger+\hat{a})).
\end{equation}
Within this approximation, the terms such as $\bar\Delta_c[3C_E^2-iC_E^nC_K/n!]$ ($n$ integer) are negligibly small. The relevant experimental values that confirm the validity of the approximation are discussed in detail in Sec.~\ref{Section7}.

In the next step, we introduce the deformed annihilation and creation operators $\hat{A}$ and $\hat{A}^\dagger$ as
\begin{equation}\label{A}
\hat{A}=(i\sqrt{C_K\hat{n}}+1)\hat{a}, \quad \hat{A}^\dagger=\hat{a}^\dagger(-i\sqrt{C_K\hat{n}}+1),
\end{equation}
in terms of which the effective Hamiltonian~\eqref{H_eff_D} can be expressed as that of a free deformed oscillator
\begin{equation}\label{HA}
\hat{H}_{\rm eff} = - \bar{\Delta}_c \hat{A}^\dagger\hat{A}.
\end{equation}
By using $[\hat{a},\hat{a}^\dagger]=1$ and $[\hat{a},\sqrt{\hat{n}}]=\frac{1}{2\sqrt{\hat{n}}}\hat{a}$, it is straightforward to show that the deformed operators $\hat{A}$ and $\hat{A}^\dagger$ obey the (deformed) commutation relation   
\begin{equation}\label{communication}
[\hat{A},\hat{A}^\dagger]=1 + C_K (2n+\frac{1}{4}e^{1/n}),
\end{equation}
which, obviously, reduces to the conventional (nondeformed) commutation relation in the absence of optomechanical coupling $(g_K=0)$.

Let us now assume that the system is manipulated through the time modulation of the driving laser frequency $\omega_L$ over the time scale $ \Gamma_m^{-1} \ll t \ll \kappa^{-1} $ according to the harmonical law, i.e.,  $ \omega_L(t)= \omega_L (1+\epsilon_L \sin\Omega t) $ with $ \epsilon_L \ll 1 $ and $\Omega$ being the amplitude and frequency of modulation, respectively. This leads to the time modulation of the detuning parameter $\bar{\Delta}_c$ according to
\begin{equation}\label{}
\bar\Delta_c \to \omega_c^{\rm eff}(t) = \Delta_c (1+\epsilon \sin\Omega t )+2g_K,
\end{equation}
where $ \epsilon = \epsilon_L \omega_L /\Delta_c$ and $\omega_c^{\rm eff}$ can be regarded as an “effective time-dependent frequency” for the primary cavity mode. Therefore, the effective Hamiltonian in Eq.~\eqref{HA} becomes time-dependent as
\begin{equation}\label{1}
\hat{H}_{\rm eff}(t)= -\omega_c^{\rm eff}(t)\hat{A}^\dagger(t)\hat{A}(t).
\end{equation}

As discussed in literature \cite{Law1994,Dodonov2011,Dodonov1996,Saito2002,Nation2012}
a single resonant cavity mode with an externally prescribed time-dependent frequency can be used as a paradigm for understanding the mechanism underlying the photon generation. Actually, harmonic time-dependence of the cavity eigenfrequency is analogous to an apparent periodic displacement of the cavity mirrors which is responsible for parametric amplification of the cavity field mode in the DCE. Within the framework of instantaneous mode functions and the associated dynamical Fock space \cite{Law1994}, the dynamics of the cavity field in the absence of dissipation can be effectively described by the Hamiltonian
\begin{equation}\label{H_DCE}
\hat{H}_{\rm DCE} = -\omega_c^{\rm eff}(t) \hat{A}^\dagger\hat{A} + i\chi(t) (\hat{A}^{\dagger 2}-\hat{A}^2),
\end{equation}
where $\hat{A}$ and $\hat{A}^\dagger$ are instantaneous operators and 
\begin{equation}
\chi(t)=\frac{1}{4{\omega}_c^{\rm eff}(t)}\frac{d{\omega}_c^{\rm eff}(t)}{dt} =\frac{\epsilon \Delta_c {\bar\Delta}_c}{2{\omega}_c^{\rm eff}(t)}  \cos(2\bar{\Delta}_c t).
\end{equation}
Here, we have taken the modulation frequency to be twice the shifted cavity detuning, $\Omega=2\bar\Delta_c$. Since $\epsilon\ll1$ we can use approximation $\omega_c^{\rm eff}\approx\bar\Delta_c$. Therefore
\begin{equation}\label{C_E}
C_\epsilon(t):=\frac{\chi(t)}{\bar\Delta_c}=\widetilde{C}_\epsilon\cos(2\bar\Delta_c t), \quad \widetilde{C}_\epsilon = \frac{\epsilon}{2}\frac{\Delta_c}{\bar\Delta_c}.
\end{equation}
The second term in Hamiltonian~\eqref{H_DCE} describes the parametric amplification of the vacuum fluctuation of the cavity field which is responsible for the DCE.

Inserting Eq.~\eqref{A} into Eq.~\eqref{H_DCE} and then using the inverse transformation of Eq.~\eqref{H_eff_D} we obtain
\begin{eqnarray}\label{H_DCE_total}
\hat{H}_{\rm DCE}&&=\hat{D}^\dagger\hat{\tilde{H}}_{\rm DCE}\hat{D}=
\nonumber
\\
&&-\bar\Delta_c\Big[\hat a^\dag \hat a  +  C_K \hat a^{\dag 2} \hat a^2 - iC_E(\hat{a}^\dagger-\hat{a})-iC_\epsilon(t) (\hat{a}^{\dagger 2}-\hat{a}^2)
\nonumber
\\
&&-C_\epsilon(t)\sqrt{C_K}(\hat{a}^{\dagger 2}\sqrt{\hat{n}}+\hat{a}^\dagger\sqrt{\hat{n}}\hat{a}^\dagger+\hat{a}\sqrt{\hat{n}}\hat{a}+\sqrt{\hat{n}}\hat{a}^2)
\nonumber
\\
&&+iC_\epsilon(t) C_K(\hat{a}^\dagger\sqrt{\hat{n}}\hat{a}^\dagger\sqrt{\hat{n}}-\sqrt{\hat{n}}\hat{a}\sqrt{\hat{n}}\hat{a})
\Big]
\nonumber
\\
&&-\bar\Delta_cC_E^2
-\bar\Delta_c \sum_{n=1}\frac{C_E^n}{n!}\big[
C_K\hat a^{\dag 2} \hat a^2-iC_\epsilon(t) (\hat{a}^{\dagger 2}-\hat{a}^2)
\nonumber
\\
&&-C_\epsilon(t)\sqrt{C_K}(\hat{a}^{\dagger 2}\sqrt{\hat{n}}+\hat{a}^\dagger\sqrt{\hat{n}}\hat{a}^\dagger+\hat{a}\sqrt{\hat{n}}\hat{a}+\sqrt{\hat{n}}\hat{a}^2)
\nonumber
\\
&&+iC_\epsilon(t) C_K(\hat{a}^\dagger\sqrt{\hat{n}}\hat{a}^\dagger\sqrt{\hat{n}}-\sqrt{\hat{n}}\hat{a}\sqrt{\hat{n}}\hat{a})
\big].
\end{eqnarray}
Finding an analytical solution for the system dynamics with such a complicated Hamiltonian is a task so difficult that it is not worth the effort. However, certain physical approximations can be made in order to simplify the Hamiltonian of Eq.~\eqref{H_DCE_total}. In the dispersive limit $(\Delta_c\gg1)$ and for weak modulation amplitude $(E_L\ll1)$ we have $C_E\ll1$(see first relation in Eq.~\eqref{C}), and we can approximate the Hamiltonian of Eq.~\eqref{H_DCE_total} to
\begin{eqnarray}\label{H_DCE2}
\hat{H}_{\rm DCE}\approx-\bar\Delta_c
\Big[
&&\hat a^\dag \hat a + C_K \hat a^{\dag 2} \hat a^2 - i C_E(\hat a^\dag-\hat a) - iC_\epsilon(t)(\hat a^{\dag 2} - \hat a^2)
\nonumber
\\
&&-C_\epsilon(t) \sqrt{C_K}(\hat{a}^{\dagger 2}\sqrt{\hat{n}}+\hat{a}^\dagger\sqrt{\hat{n}}\hat{a}^\dagger+\hat{a}\sqrt{\hat{n}}\hat{a}+\sqrt{\hat{n}}\hat{a}^2)
\nonumber
\\
&&+iC_\epsilon(t) C_K(\hat{a}^\dagger\sqrt{\hat{n}}\hat{a}^\dagger\sqrt{\hat{n}}-\sqrt{\hat{n}}\hat{a}\sqrt{\hat{n}}\hat{a})
\Big].
\end{eqnarray}
If we additionally assume the optomechanical coupling to be weak $(g_K\ll1)$ then $C_K\ll1$ (see second relation in Eq.~\eqref{C}).
In this case, which we refer to as the \textit{weak coupling regime} (WCR), one can approximately drop the terms proportional to $C_\epsilon\sqrt{C_K}$ and $C_\epsilon C_K$ in Eq.~\eqref{H_DCE2} to obtain a more simplified Hamiltonian as
\begin{equation} \label{H_WR}
\hat{H}_{\rm WCR}=-\bar\Delta_c\Big[
\hat a^\dag \hat a + C_K \hat a^{\dag 2} \hat a^2 - i C_E(\hat a^\dag-\hat a)-iC_\epsilon(t)(\hat a^{\dag 2} - \hat a^2)\Big].
\end{equation}
We address the time evolution of the system in a twofold manner. On the one hand, we explore the dissipative dynamics by numerically simulating the master equation based on both Hamiltonians~\eqref{H_DCE2}~and~\eqref{H_WR}. On the other hand, it is important to note that the WCR Hamiltonian—unlike the more general Hamiltonian~\eqref{H_DCE2}—admits an approximate analytical solution. Specifically, by considering the WCR regime and neglecting dissipation, we can derive a closed-form expression for the system's time-evolution operator. This analytical insight not only provides an independent perspective on the system’s behavior but also serves as a valuable benchmark for evaluating the accuracy and robustness of the numerical simulations.

\subsection{Numerical solution}
To simulate the system dynamics by solving its governing master equation numerically, we assume that the cavity field is coupled to a vacuum reservoir. Considering the Markovian dissipation, the evolution of the system is governed by the quantum master equation
\begin{eqnarray}\label{master-equation}
\dot{\hat{\rho}}=&&i[\hat{\rho},\hat{H}_{\rm DCE (WCR)}]+\frac{\kappa}{2}(\hat{a}\hat{\rho}\hat{a}^\dagger-\hat{a}^\dagger\hat{a}\hat{\rho}-\hat{\rho}\hat{a}^\dagger\hat{a}),
\end{eqnarray}
where the Hamiltonians $\hat{H}_{\rm DCE (WCR)}$ are given in Eqs.~\eqref{H_DCE2}, and~\eqref{H_WR}, respectively.
By using the Quantum Toolbox in Python (QuTip) \cite{qutip1,qutip2}, we numerically solve the master equation to obtain the time-evolved density matrix of the system  $\hat{\rho}(t)$ by which we can evaluate the mean number of the generated microwave Casimir photons and their quantum statistical properties.
Note that solving the master equation with Hamiltonian \eqref{H_DCE2} enables us to obtain the exact dynamics of the system beyond the WCR. 

It is important to mention that our numerical analysis must rely on the effective Hamiltonians in Eqs.~\eqref{H_DCE2} and~\eqref{H_WR}, as a direct simulation of the optomechanical Hamiltonian in Eq.~\eqref{H2} is not practically feasible. 
In the instantaneous-mode quantization required for time-dependent cavity boundaries, the operators and the effective detuning acquire explicit time dependence, making the operator basis itself time-dependent and unsuitable for standard numerical methods. Applying the canonical transformation of Law’s dynamical-Casimir formalism \cite{Law1994} removes this difficulty by mapping the instantaneous operators to a fixed mode and transferring all time dependence to the Hamiltonian parameters. This yields the effective Hamiltonian in Eq.~\eqref{1}, from which Eqs.~\eqref{H_DCE2} and~\eqref{H_WR} follow and which provide the only numerically tractable formulation for our simulations.

\subsection{Analytical solution}

Leaving the details of derivation to appendix~\ref{Appendix_B} for simplicity of presentation, we here quote the final expression for the time evolution operator corresponding to the Hamiltonian of Eq.~\eqref{H_WR} in the absence of dissipation
\begin{eqnarray}\label{u_final}
\mathcal{\hat U}(t)=&& 
{\rm exp} \left[ \frac{\beta(t)}{4}-i\frac{g_K}{2} t\right] {\rm exp}\left[-it( \bar{\Delta}_c \hat n + g_K \hat n^2) \right]
\nonumber
\\
&&\times{\rm exp} \left[ \frac{\alpha(t)}{2} \hat a^{\dag 2} \right] {\rm exp} \left[ \frac{\beta(t)}{2} \hat n \right] {\rm exp} \left[ \frac{\gamma(t)}{2} \hat a^{2} \right],
\end{eqnarray}
with
\begin{subequations}\label{dif-ans}
	\begin{equation}\label{alpha}
	\alpha(t)=\frac{2\chi e^{4ig_K t } \sinh(\mathcal{G} t)}{\mathcal{G} \cosh (\mathcal{G} t) + i g_K \sinh(\mathcal{G}t)}, 
	\end{equation}
	\begin{equation} \label{beta}
	\beta(t)= 4ig_K t + 2\ln \mathcal{G} - 2 \ln [\mathcal{G} \cosh(\mathcal{G}t) + ig_K \sinh (\mathcal{G}t)],
	\end{equation}
	\begin{equation}\label{gamma}
	\gamma(t)=\frac{-2 \chi\sinh(\mathcal{G} t)}{\mathcal{G}\cosh (\mathcal{G} t) + i g_K\sinh(\mathcal{G}t) },
	\end{equation}
\end{subequations}
where $ \mathcal{G}:= \sqrt{ 4\chi'^2 -g_K^2 }$ is an effective coupling strength which for $ g_K > 2\chi'$ with $\chi'=\frac{\widetilde{C}_\epsilon}{2}\bar{\Delta}_c$, becomes pure imaginary, $ \mathcal{G} \to \mathcal{\bar G}=i \mathcal{G}$. 
Equation~\eqref{u_final} implies that the dynamical behaviour of the system depends on the value of $\mathcal{G}$ which is controllable through the modulation depth $\epsilon$ as well as the detuning parameter $\Delta_c$ [see Eq.~\eqref{C_E}]. This feature of the present scheme, in particular, is advantageous in the context of controlling system dynamics by external parameters.


\section{Casimir Photon Generation}\label{Section3}
In this section, we investigate the possibility of generating photons due to the DCE within the framework developed in~Sec.\ref{Section2}. To this end, we consider two distinct Hamiltonians: the full effective Hamiltonian given in Eq.\eqref{H_DCE2}, which is treated numerically, and its simplified counterpart corresponding to the WCR, presented in Eq.\eqref{H_WR}, which is examined both numerically and analytically.

\begin{figure}
	\centering
	\includegraphics[width=\linewidth]{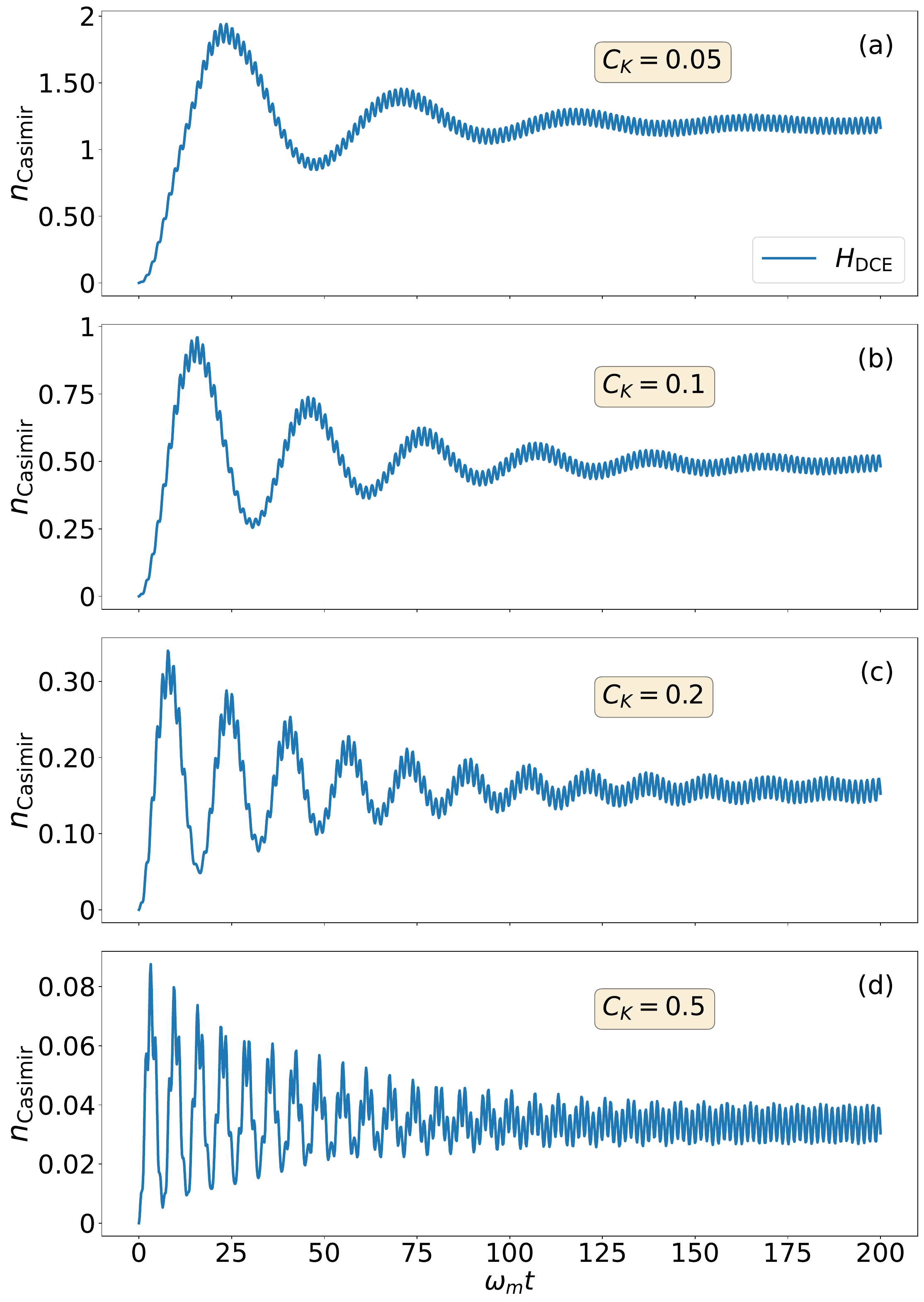}
	\caption{
	Time evolution of the mean number of generated Casimir photons $n_{\rm Casimir} =\bra{0}\hat{n}(t)\ket{0}$ based on the exact numerical solution of the master equation~\eqref{master-equation} with Hamiltonian $\hat{H}_{\rm DCE} $ in Eq.~\eqref{H_DCE2} for different values of the scaled Kerr nonlinearity parameter $C_K$.
	The other system parameters are set as $\omega_m/2\pi=5.33 \text{MHz}$, $\bar{\Delta}_c=\omega_m$, $\widetilde{C}_\epsilon=10^{-1}$, $C_E=10^{-2}$, $\kappa/2\pi=118 \text{kHz}$. 
	}
	\label{fig2}
\end{figure}

Numerical results corresponding to the full effective Hamiltonian of~\eqref{H_DCE2} are presented in Fig.~\ref{fig2}, which illustrates the time evolution of the mean number of generated Casimir photons. 
As seen, for $C_K < \widetilde{C}_\epsilon$ [Fig.~\ref{fig2}(a)] the mean number of Casimir photons increases in the early stages of the system evolution followed by rapidly damped oscillations, and eventually, it relaxes to a stationary value. As $C_K$ is increased such that it exceeds $\widetilde{C}_\epsilon(C_K>\widetilde{C}_\epsilon)$, we see from Figs.~\ref{fig2}(b)-(d) that the oscillatory behavior of the mean number of Casimir photons becomes more obvious and prominent. 
Moreover, lower values of the Kerr coefficient can lead to a higher peak in the number of generated photons, highlighting the nontrivial influence of the Kerr nonlinearity on photon generation dynamics.

\begin{figure}
	\centering
	\includegraphics[width=\linewidth]{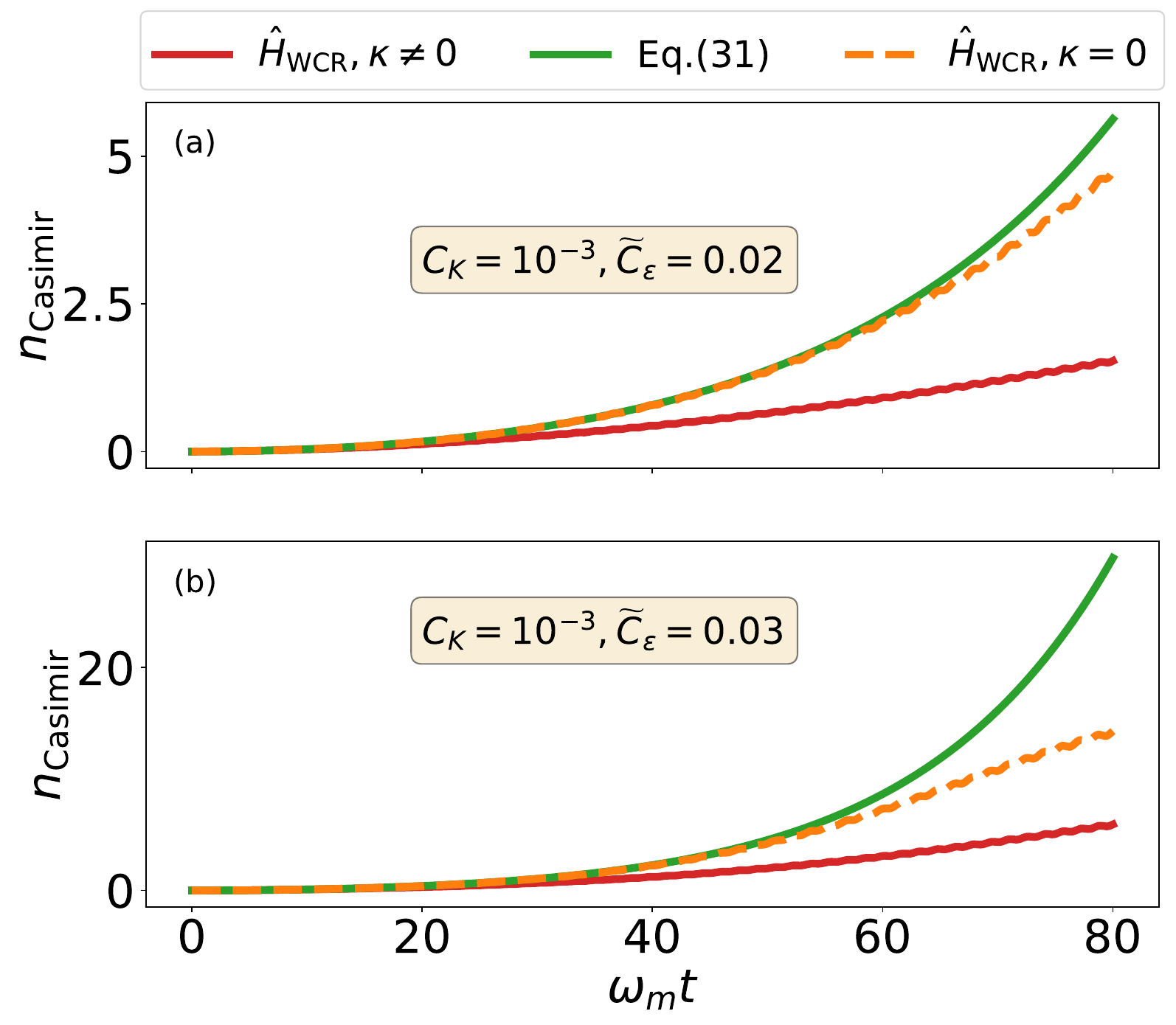}
	\caption{
Time evolution of the mean number of generated Casimir photons 
$n_{\rm Casimir}(t) = \langle 0 | \hat{n}(t) | 0 \rangle$ in the WCR, obtained from both the analytical expression in Eq.~\eqref{n_Casimir} and numerical simulations based on Hamiltonian~\eqref{H_WR}.
The orange dashed line shows the evolution in the absence of dissipation.
The system parameters are set as $\omega_m/2\pi = 5.33~\text{MHz}$, $\bar{\Delta}_c = \omega_m$, $C_E = 10^{-2}$, and $\kappa/2\pi = 118~\text{kHz}$.
Here, the Kerr nonlinearity strength is fixed to $C_K=10^{-3}$ while $\widetilde{C}_\epsilon$ is varied to explore its influence on photon generation.
}
	\label{fig3}
\end{figure}

We now analyze the temporal behavior of the mean number of Casimir photons in the WCR. For this purpose, we calculate analytically $n_{\rm Casimir}(t)$ by utilizing the time evolution operator given in Eq.~\eqref{u_final}. As shown in Appendix~\ref{Appendix-C}, the mean number of Casimir photons in the WCR and in the absence of cavity dissipation $(\kappa=0)$ is given by
\begin{equation}\label{n_Casimir}
n_{\rm Casimir}=\frac{4\chi'^2}{\mathcal{G}^2} \sinh^2 \mathcal{G}t.
\end{equation}

In Fig.~\ref{fig3}, we have plotted the temporal behavior of the mean number of generated Casimir photons obtained from both the analytical expression in Eq.~\eqref{n_Casimir} and the numerical simulation of the master equation~\eqref{master-equation} with the Hamiltonian $\hat{H}_{\rm WCR}$ in Eq.~\eqref{H_WR}. The results demonstrate that the analytical approximate solution (green-solid line) is in close agreement with the numerical solution in the absence of dissipation (orange-dashed line). However, when the cavity dissipation is taken into account (red-solid line), the exact numerical solution deviates from the two former solutions as time elapses.
The results also demonstrate that, for a fixed Kerr coefficient $C_K$, increasing the modulation amplitude parameter $\widetilde{C}_\epsilon$ leads to a significant enhancement in the number of generated photons. 
These observations confirm that the simplified model in the WCR accurately captures the key dynamical features of Casimir photon generation under weak coupling conditions.

The exponential growth observed in Fig.~\ref{fig3} can be described by Eq.~\eqref{n_Casimir}. When $C_K < \widetilde C_\epsilon$ (which directly implies \( g_K < 2\chi'(t) \)), the effective coupling parameter \( \mathcal{G} \) remains real, and the photon number evolves as \( n(t) \propto \sinh^2(\mathcal{G}t) \). 
This hyperbolic growth is a hallmark of the WCR and is accurately captured by the analytical solution. 
Although Eq.~\eqref{n_Casimir} is strictly valid in the WCR, its structure also helps to qualitatively explain the periodic behavior observed in Fig.~\ref{fig2}. 
In this case, when $C_K > \widetilde C_\epsilon$ (implying \( g_K > 2\chi'(t) \)), the coupling parameter becomes imaginary, \( \mathcal{G} \to i\mathcal{G} \), leading to an oscillatory behavior of the form \( n(t) \propto \sin^2(\mathcal{G}t) \). 
Obviously, it is one the advantage of our proposal in the sense that the mean number of generated Casimir photons and its dynamical behavior (oscillatory or exponential) can be controlled externally by the relative strength of the modulation amplitude and the optomechanically induced Kerr nonlinearity.

It should be noted that the Kerr term proportional to $\hat a^{\dag 2} \hat a^2$ does not have any role in the generation of photons since it conserves the number operator. 
In the other words, photon generation is only due to the DCE process through coherent time modulation. 
On the other hand, the Kerr nonlinearity plays a distinct role in this system by limiting the photon generation process even in the absence of dissipation and loss, or by inducing an oscillatory signature in the dynamics of the mean number of Casimir photons.

In order to see more clearly the role played in the Casimir photon creation by the time modulation of the driving laser frequency and the induced Kerr nonlinearity, we consider the rate of Casimir photon production. From the perspective of practical applications, notably in free-space or in fiber, the photon production rate is an important quantity which can be experimentally measured by using single-photon avalanche detectors (SPADs) \cite{vezzoli2019}. In the system under investigation, the average rates of the Casimir photon production corresponding to Figs.~\ref{fig2}(a), \ref{fig2}(b), \ref{fig3}(a), and \ref{fig3}(b) are, respectively, obtained as 10Mcps, 50Mcps, 40Mcps, and 1.23Gcps (count or photon per second)  . Accordingly, we understand that when the induced Kerr nonlinearity is extremely weak, the rate of Casimir photon creation increases significantly as the modulation amplitude parameter increases a little bit.
The high generation rate in range from MHz to GHz implies that the proposed scheme can be seen as a \textit{high-brightness nonclassical microwave quantum source}. 
(The nonclassicality of the generated Casimir radiation will be verified in the following Secs.~\ref{Section4},~\ref{Section5}, and~\ref{Section6}.)


\section{Photon Counting Statistics} \label{Section4}
\begin{figure}
	\centering
	\includegraphics[width=\linewidth]{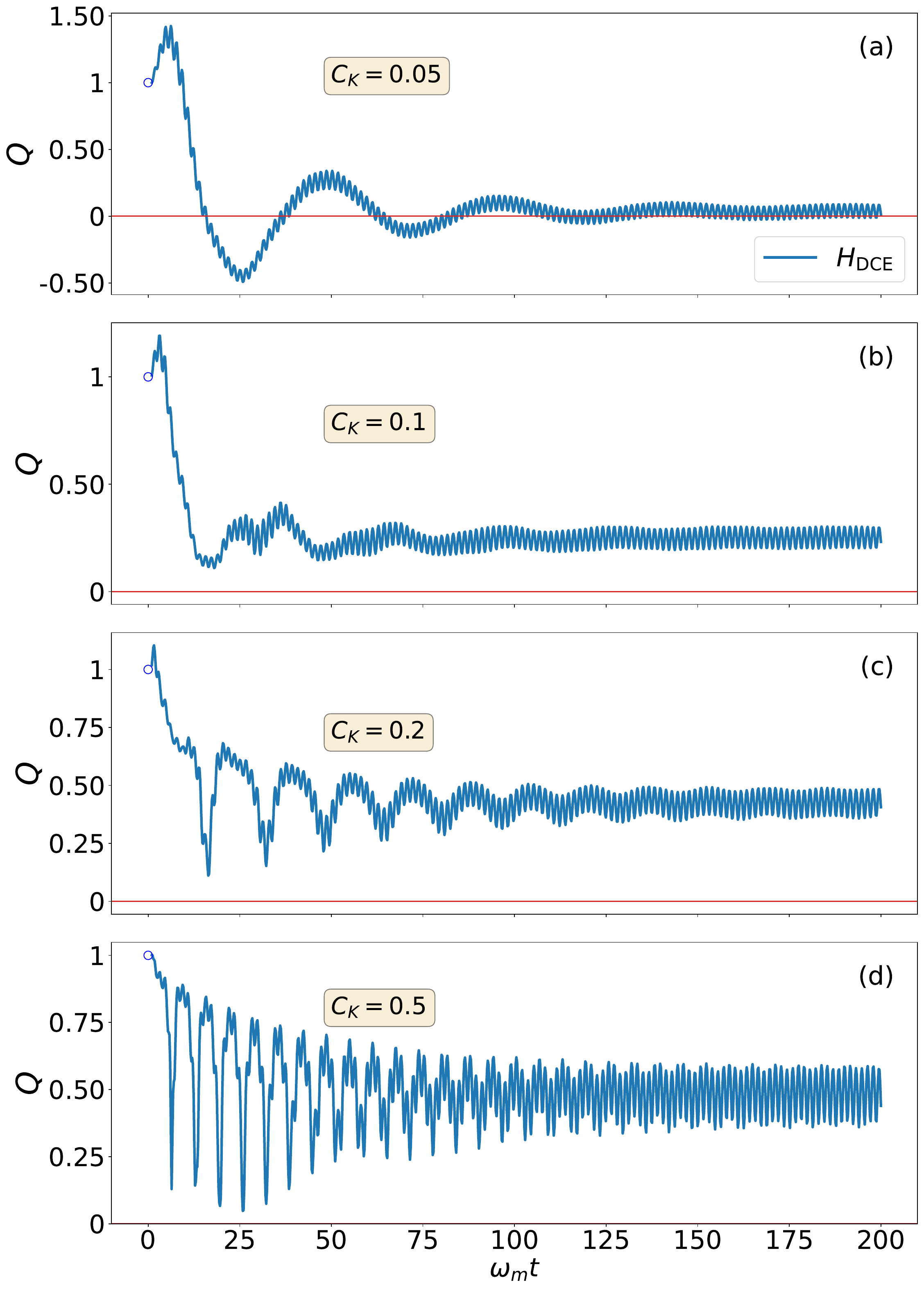}
	\caption{
	Time evolution of the Mandel parameter obtain by the exact numerical solution of master equation~\eqref{master-equation} with the Hamiltonian $\hat{H}_{\rm DCE} $ in Eq.~\eqref{H_DCE2} for different values of the scaled Kerr nonlinearity parameter.
	The other system parameters are the same as in Fig.~\ref{fig2}.
	}
	\label{fig4}
\end{figure}
\begin{figure}
	\centering
	\includegraphics[width=\linewidth]{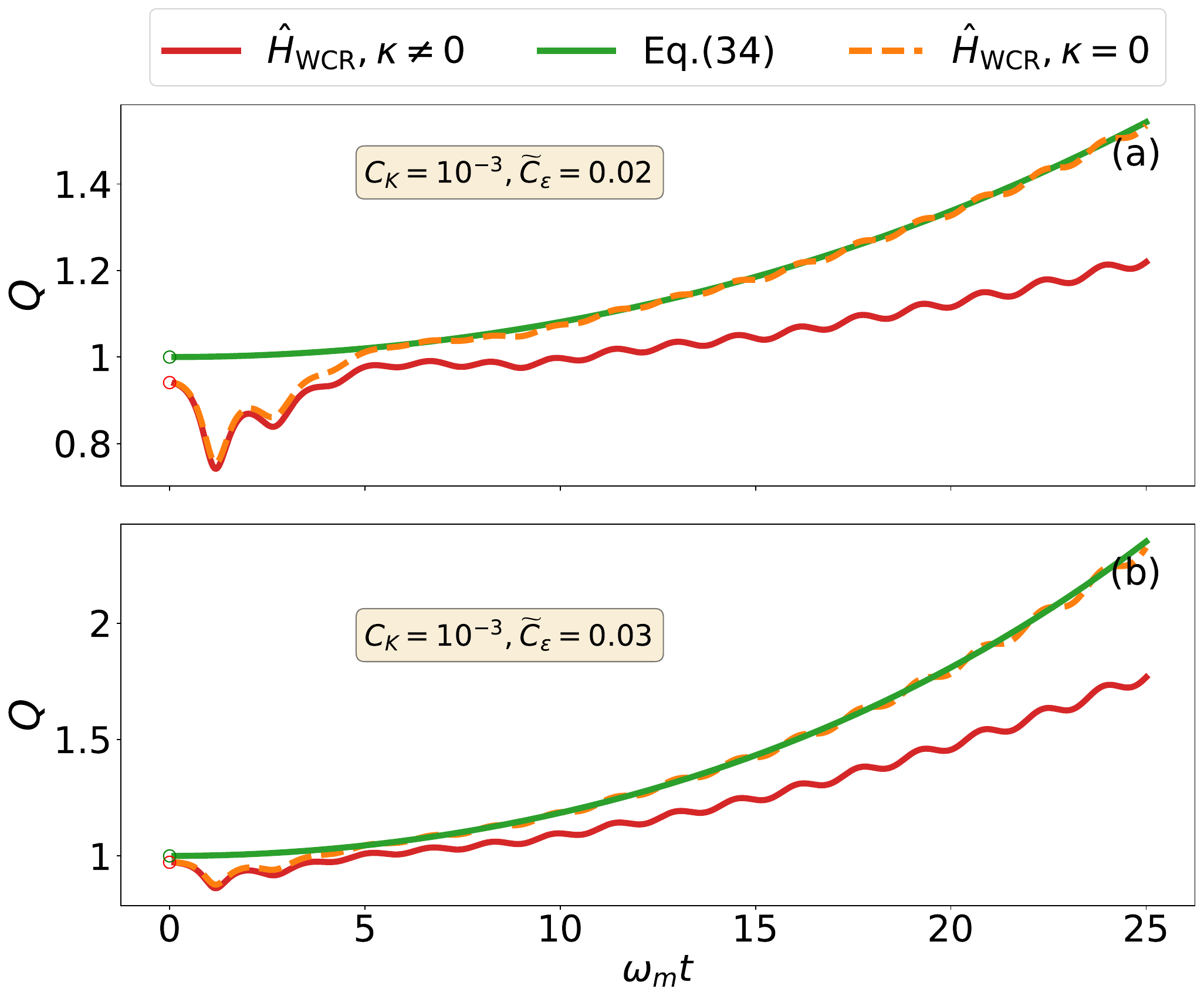}
	\caption{
Time evolution of the Mandel parameter in the WCR, obtained from both the analytical expression in Eq.~\eqref{Mandel-analytical} and numerical simulations based on Hamiltonian~\eqref{H_WR}.
The orange dashed line shows the evolution in the absence of dissipation.
The system parameters are set as $\omega_m/2\pi = 5.33~\text{MHz}$, $\bar{\Delta}_c = \omega_m$, $C_E = 10^{-2}$, and $\kappa/2\pi = 118~\text{kHz}$, while $\widetilde{C}_\epsilon$ is varied to explore its influence on photon counting statistics.
	}
	\label{fig5}
\end{figure}
We now focus on the quantum statistical properties of the generated Casimir photons, particularly their photon counting statistics as described by the Mandel parameter. This quantity indicates whether the photon statistics is sub-Poissonian, Poissonian, or super-Poissonian, and thus provides useful insight into the quantum or classical nature of the emitted radiation. The Mandel parameter is defined as~\cite{mandel1986}
\begin{eqnarray} \label{Q}
Q(t) = \frac{\langle \hat a^{\dag 2}(t) \hat a^2(t) \rangle - \langle \hat n(t) \rangle^2 }{\langle \hat n(t) \rangle}=\frac{(\Delta \hat n)^2 - \langle \hat n(t) \rangle}{\langle \hat n(t) \rangle},
\end{eqnarray}
where $ \Delta n=\sqrt{\langle \hat n^2 \rangle-\langle \hat n \rangle^2} $. For $Q>0$ ($Q<0$), the statistics is said to be super-Poissonian (sub-Poissonian); $Q=0$ stands for Poissonian statistics.

Figure~\ref{fig4} illustrates the time evolution of the Mandel parameter $Q(t)$ of the generated Casimir photons obtained by solving numerically the master equation~\eqref{master-equation} with the Hamiltonian $\hat{H}_{\rm DCE}$ in Eq.~\eqref{H_DCE2} for different values of the scaled Kerr nonlinearity parameter. 
Note that the Mandel parameter is not defined for the vacuum state as the initial state of the system, so we have marked this point with an open circle in the figure.

As can be seen from Fig.~\ref{fig4}(a), for $C_K<\widetilde{C}_\epsilon$  , the Mandel parameter decreases towards negative values and reaches $Q = -0.5$ in the early stages of the system evolution, indicating the nonclassical sub-Poissonian statistics of the generated Casimir photons. Then, after making damped oscillations, it is finally stabilized at an asymptotical zero value corresponding to the Poissonian statistics. On the other hand, Figs.~\ref{fig4}(b)-4(c) reveal that as $C_K$  increases towards values exceeding $\widetilde{C}_\epsilon$, the Mandel parameter remains positive at all times which corresponds to the classical super-Poissonian statistics.

We now examine the time evolution of the photon statistics of the generated Casimir radiation in the WCR. By using Eq.~\eqref{n1}, the variance in the photon number is obtained as 
\begin{align}\label{Var_n-cogherent}
(\Delta \hat{n})^2= \bra{0} \hat n^2(t) \ket{0}-\bra{0} \hat n(t) \ket{0}^2 =2\abs{\Phi_1}^2,
\end{align}
where $\Phi_1$ is given by Eq.~\eqref{phi1}.
Therefore, the Mandel parameter for $t>0$ is
\begin{equation}\label{Mandel-analytical}
Q(t) = 1+ \frac{8\chi'^2}{\mathcal{G}^2}\sinh^2(\mathcal{G}t).
\end{equation}

In Fig.~\ref{fig5}, we have depicted the Mandel parameter for the generated Casimir photons as a function of time, obtained from both the analytical expression in Eq.~\eqref{Mandel-analytical} and the numerical simulation of the master equation Eq.~\eqref{master-equation} with the Hamiltonian  $\hat{H}_{\rm WCR}$ in Eq.~\eqref{H_WR} for both cases of the presence and the absence of cavity dissipation. 
The results demonstrate that in the WCR and when $C_K$ is extremely small, the Mandel parameter grows monotonically such that the generated Casimir photons always obey the super-Poissonian statistics. Moreover, in the absence of cavity dissipation, there is a good agreement between the numerical solution (green-solid line) and the  analytical solution (orange-dashed line) as time goes on. In the realistic situation, where the cavity dissipation is taken into account (red-solid line), the Mandel parameter increases in a slower manner as compared with the case of no cavity dissipation.

\section{quadrature squeezing}\label{Section5}
In view of the generation mechanism of quadrature squeezed states in one hand, and the presence of the term proportional to $\hat{a}^{2 \dagger}-\hat{a}^2$ in the Hamiltonians of Eqs.~\eqref{H_DCE2} and~\eqref{H_WR} on the other hand, it is reasonable to expect that the generated Casimir photons exhibit the squeezing effect. 
To explore the squeezing property of the generated Casimir photons, we define the dimensionless quadrature amplitudes $\hat{q}=\frac{\hat{a}+\hat{a}^\dagger}{2}$ and $\hat{p}=\frac{\hat{a}-\hat{a}^\dagger}{2i}$ related to the position and momentum operators of the cavity mode, respectively. A quantum state of the cavity mode is said to be squeezed when one of the quadrature components $\hat{q}$ and $\hat{p}$ satisfies the relation $\expval{(\delta \hat{O})^2}<1/4(\hat{O}=\hat{q}\text{\space or\space}\hat{p})$. The degree of quadrature squeezing can be quantified in dB (decibel) unit via 
\begin{equation}\label{Def-Sq}
S_O=- 10 \log_{10}\frac{\expval{(\delta \hat{O})^2}}{\expval{(\delta \hat{O})^2}_{\rm vac}}, \qquad (O=q,p)
\end{equation}
with $\expval{(\delta \hat{O})^2}_{\rm vac}=1/4$ as the quantum-vacuum fluctuation. Then, the condition for squeezing in the quadrature component can be simply written as $S_O>0$. Since 3-dB squeezing, which corresponds to $50\%$ noise reduction below the zero-point level (i.e., $\expval{(\delta \hat{O})^2}=\expval{(\delta \hat{O})^2}_{\rm vac}/2=1/8$) is a limit for many proposals, the squeezing beyond 3 dB can be regarded as strong squeezing. In the WCR, by using the Hamiltonian~\eqref{H_DCE2}, the variances of the cavity mode quadratures fluctuations can be obtained as (see Appendix~\ref{Appendix_Squeezing}).

\begin{subequations}\label{var_q_p}
\begin{equation}\label{var-q}
(\Delta \hat{q})^2= \frac{1 + 2 \Phi_4(t)- (\mu(t) \nu(t) + \lambda(t) \kappa(t)) }{4},
\end{equation}
\begin{equation}\label{var-p}
(\Delta \hat{p})^2= \frac{1 -  2 \Phi_4(t) + (\mu(t) \nu(t) + \lambda(t) \kappa(t))}{4},
\end{equation}
\end{subequations}
where the parameters $\mu,\nu,\lambda,$ and $\kappa$ are defined in Eqs.~\eqref{Coefficient} and $\Phi_4$ is given by Eq.~\eqref{n_casimir_app}.

\begin{figure}
	\centering
	\includegraphics[width=\linewidth]{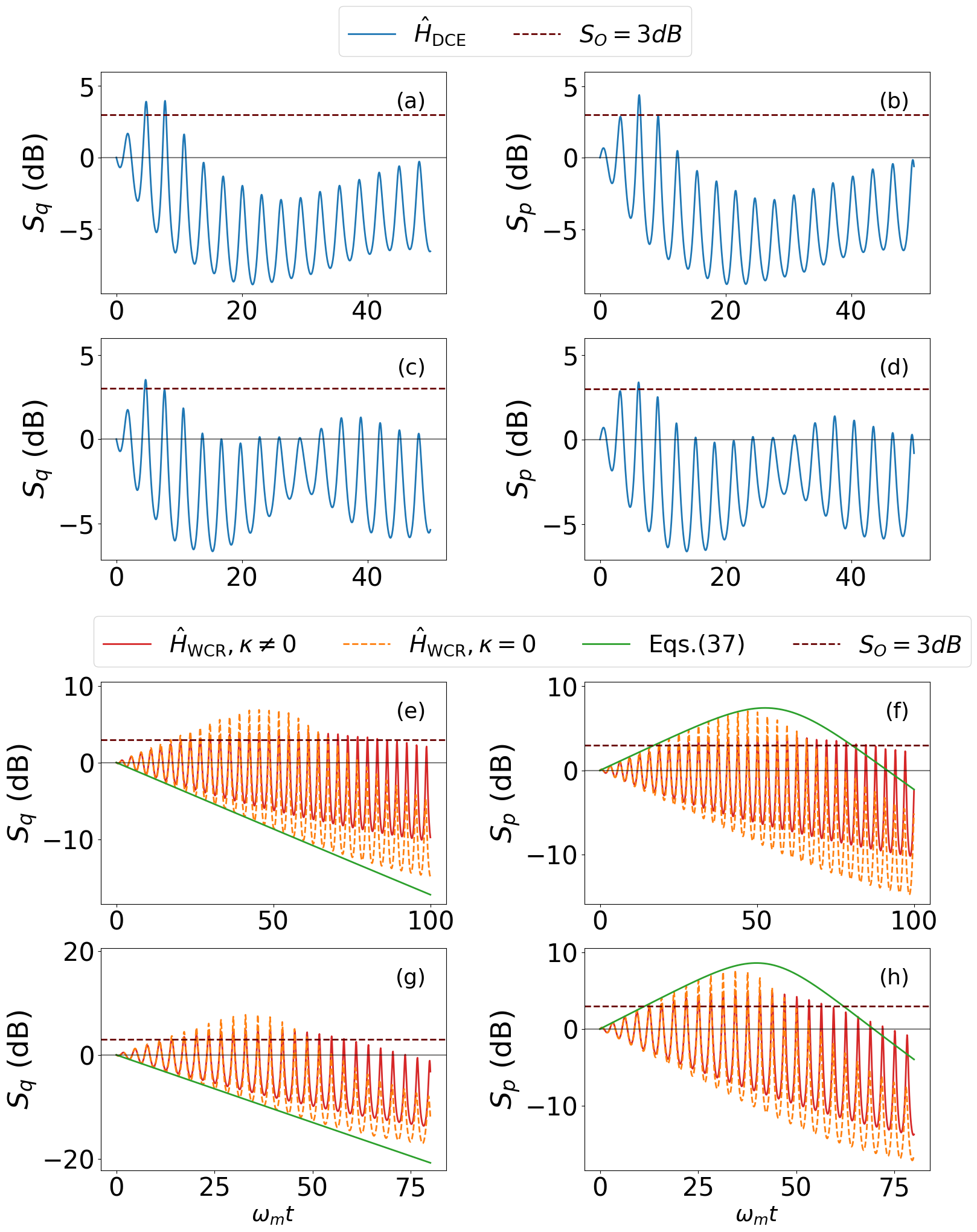}
	\caption{
 Time evolution of $S_q(t)$ (left panels) and $S_p(t)$ (right panels) for the generated DCE photons. Panels (a)-(d) show the results obtained from the numerical solution of the master equation~\eqref{master-equation} with the Hamiltonian $\hat{H}_{\rm DCE}$ [Eq.~\eqref{H_DCE2}] for the fixed value of $\widetilde{C}_\epsilon=0.1$ and for two values of $C_K$: $C_K=0.05$ (panels a and b) and $C_K=0.1$ (panels c and d). Panels (e)-(h) show the results obtained using the analytical expression in Eqs.~\eqref{var_q_p} (green-solid curves) as well as the numerical simulation of the master equation~\eqref{master-equation} with the Hamiltonian $\hat{H}_{\rm WCR}$, with and without considering the cavity dissipation (red-solid and orange-dashed curves, respectively). In these panels we have fixed $C_K=10^{-3}$ and have varied $\widetilde{C}_\epsilon$: $\widetilde{C}_\epsilon=0.02$(panels e and f) and $\widetilde{C}_\epsilon=0.03$ (panels g and h). 
 The other system parameters are the same as in Fig.~\ref{fig2}.
	}
	\label{fig6}
\end{figure}

Now, we examine the temporal behaviour of $S_O(t)$ which gives information on the squeezing of the quadrature $O(O=q,p)$. In Figs.~\ref{fig6}(a)-(d) we have plotted the squeezing parameters $S_q(t)$ (left panels) and $S_p(t)$ (right panels) for the generated Casimir photons with respect to time, obtained from the numerical solution of the master equation~\eqref{master-equation} with the Hamiltonian  [Eq.~\eqref{H_DCE2}]. For panels (a) and (b) we take the Kerr coefficient $C_K=0.05$ and for panels (c) and (d) we choose $C_K=0.1$. 
The modulation amplitude parameter $\widetilde{C}_\epsilon$ is fixed at value $0.1$. It is seen that both $S_q(t)$ and $S_p(t)$ oscillate as functions of time, showing alternatively some degree of quadrature squeezing (up to 4.3 dB) at short times, but any squeezing disappears as time passes due to the cavity field dissipation. We also find that with increasing the Kerr coefficient, the quadratures squeezing can reappear at longer times, albeit with smaller amplitudes. Figs.~\ref{fig6}(e)-(h) demonstrate the time evolution of $S_q(t)$ (left panels) and $S_p(t)$ (right panels) calculated by using the analytical expressions in Eqs.~\eqref{var_q_p} (green-solid curves)  as well as the numerical simulation of the master equation~\eqref{master-equation} with the Hamiltonian $\hat{H}_{\rm WCR}$ in Eq.~\eqref{H_WR}, with and without considering the cavity dissipation (red-solid and orange-dashed curves, respectively). As can be seen, in the WCR and with the same cavity dissipation rate as used in the numerical analysis with the Hamiltonian $\hat{H}_{\rm DCE}$ (red-solid curves), a higher quadrature squeezing (up to 5dB) can be achieved.
More importantly, the quadratures squeezing can persist over a relatively longer time, indicating an improved resilience to dissipation. These results suggest that the WCR described by Eq.~\eqref{H_WR} provides better performance and robustness against environmental losses compared to the full effective Hamiltonian $\hat{H}_{\rm DCE}$ in Eq.~\eqref{H_DCE2}.

Also the green-solid curves in Fig.~\ref{fig6} are corresponding the analytical solutions of Eqs.~\eqref{var_q_p}.
The analytical solutions closely resemble the envelope of the numerical results in Figs.~\ref{fig6}(e,f,g,h). 
This similarity arises from certain approximations applied to derive the analytical equation (see Appendix~\ref{Appendix_B}), which effectively suppress the oscillatory behaviors emerging in the numerical results.

\section{Nonclassicality of the generated DCE Photons in Phase Space}\label{Section6}

Beyond the investigation of the Mandel parameter and quadrature squeezing, the generation of nonclassical states emerges as a noteworthy quantum feature of the proposed optomechanical system. In what follows, we examine the potential for generating nonclassical states of Casimir photons enabled by the intrinsic Kerr-type optomechanical nonlinearity inherent in the system operating within the RDR.
This capability is fundamentally linked to the presence of the Kerr interaction term $ g_k \hat a^{\dagger 2} \hat a^2 $, which is known to facilitate the generation of Yurke-Stoler states when the system evolves from an initial coherent state over a characteristic time scale $ \tau = \pi /2g_k $. Notably, coherent states themselves can be efficiently prepared from an initial vacuum state via displacement-type interactions such as $ C_E (\hat a^\dagger - \hat a) $. Since the full system Hamiltonian contains several additional interaction terms beyond the Kerr nonlinearity, it is reasonable to expect that the exact generation of a Yurke-Stoler cat state may be hindered. However, the interplay among these terms can still give rise to approximate nonclassical states for the generated microwave Casimir photons.

The signature of the nonclassical characteristic of the generated states can be revealed by calculating their corresponding Wigner function in the position-momentum phase space which is defined as\cite{schleich} 
\begin{equation}\label{Wigner_dis}
W(x,p)=\frac{1}{2\pi}\int_{-\infty}^\infty dy \bra{x+\frac{y}{2}}\hat{\rho}\ket{x-\frac{y}{2}}\exp(-ipy),
\end{equation}
where $\hat{\rho}$ is defined in Eq.~\eqref{master-equation}, and $x$ and $p$ are the position and momentum variables in the phase space, respectively.

\begin{figure*}
	\centering
	\includegraphics[width=\linewidth]{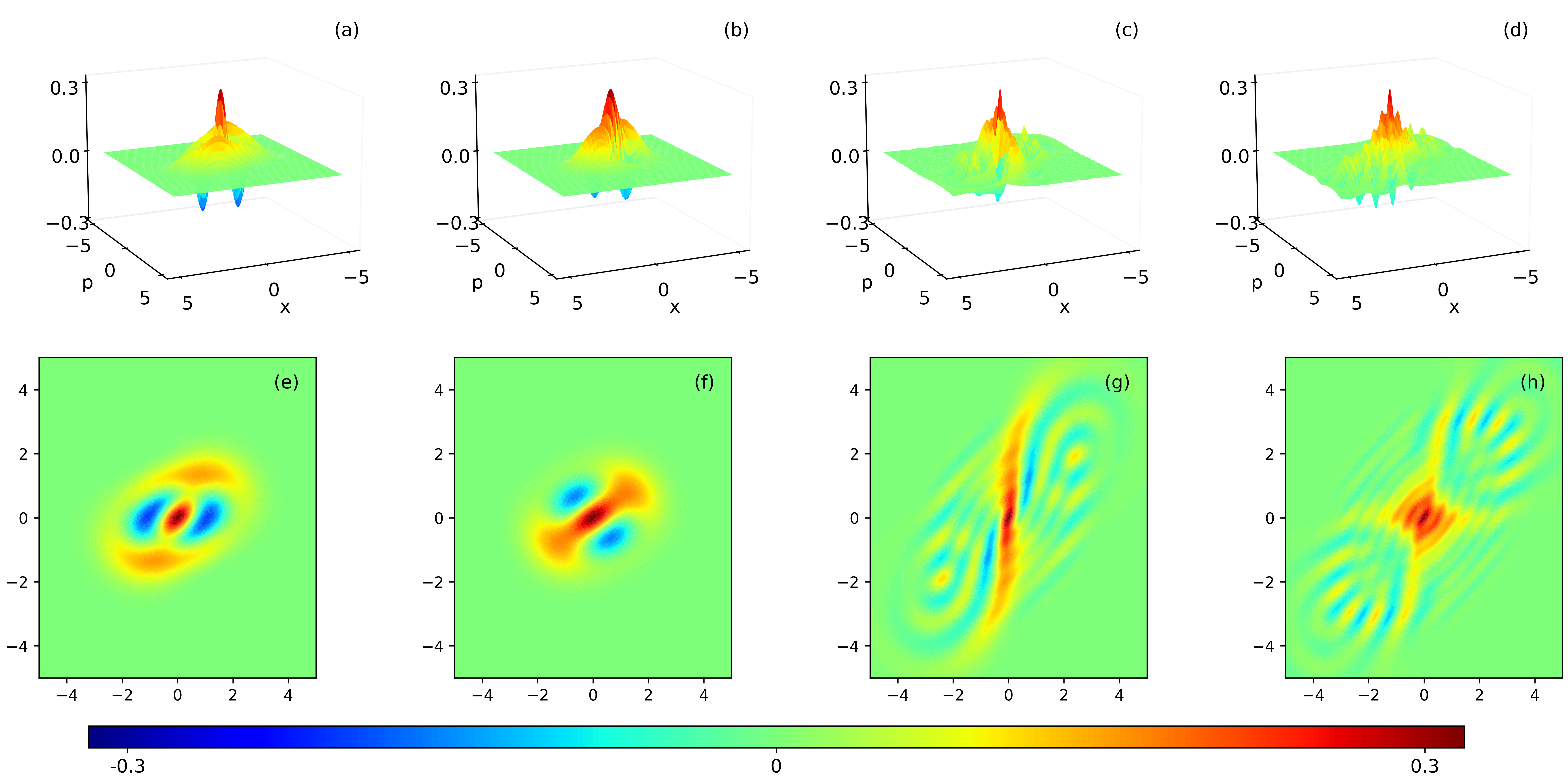}
	\caption{
Three dimensional plots of the Wigner function, $W(x,p)$, along with its corresponding contour plots, at the time scale $\tau = \pi/2g_K$, plotted versus dimensionless canonical quadratures $x$ and $p$. Each 3D surface plot in (a)–(d) has its corresponding contour plot placed directly below it in (e)–(h), respectively. The parameter sets used are: (a,e) $C_K = 0.05$, $\widetilde{C}_\epsilon = 0.1$; (b,f) $C_K = 0.1$, $\widetilde{C}_\epsilon = 0.1$; (c,g) $C_K = 10^{-3}$, $\widetilde{C}_\epsilon = 0.02$; and (d,h) $C_K = 10^{-3}$, $\widetilde{C}_\epsilon = 0.03$. Other system parameters are fixed as $\omega_m/2\pi = 5.33~\text{MHz}$, $\bar{\Delta}_c = \omega_m$, $C_E = 10^{-2}$, and $\kappa = 0$.
	}
	\label{fig7}
\end{figure*}

Figure~\ref{fig7} illustrates the Wigner function obtained by numerically solving the master equation~\eqref{master-equation} using QuTiP, accounting for dissipation. This section does not include any analytical plots. The functions are evaluated at the specific time $\tau = \pi/2g_K$, chosen based on the structure of the engineered state, which enables the formation of a Yurke-Stoler cat state within the framework of the Kerr Hamiltonian. Figures~\ref{fig7}(a) and (b) correspond to the dynamics governed by Hamiltonian~\eqref{H_DCE2}, while Figures~\ref{fig7}(c) and (d) correspond to the WCR Hamiltonian defined in Eq.~\eqref{H_WR}.

When the Kerr nonlinearity parameter \( C_k \) is relatively small in the WCR (see Figs.~\ref{fig7}(c) and (d)), the resulting quantum states fail to exhibit a well-defined superposition. In contrast, for higher values of \( C_k \) within the framework of the full effective Hamiltonian (Figs.~\ref{fig7}(a) and (b)), the photon states manifest distinctly nonclassical features in their Wigner functions, characterized by the appearance of pronounced interference peaks at time \(\tau\), in agreement with theoretical predictions. 
This comparison demonstrates that the Hamiltonian presented in Eq.~\eqref{H_DCE2} exhibits a greater capacity to generate highly nonclassical states as evident in Figs.~\ref{fig7}(a,e) and (b,f)—than the WCR Hamiltonian described in Eq.~\eqref{H_WR}, as shown in Figs.~\ref{fig7}(c,g) and (d,h). 
Consequently, a comparison between Hamiltonians~\eqref{H_DCE2} and~\eqref{H_WR} for different values of $C_K/\widetilde{C}_\epsilon$ reveals that higher values of \( g_K \) are more effective in producing strongly nonclassical states.

\begin{figure}
	\centering
	\includegraphics[width=\linewidth]{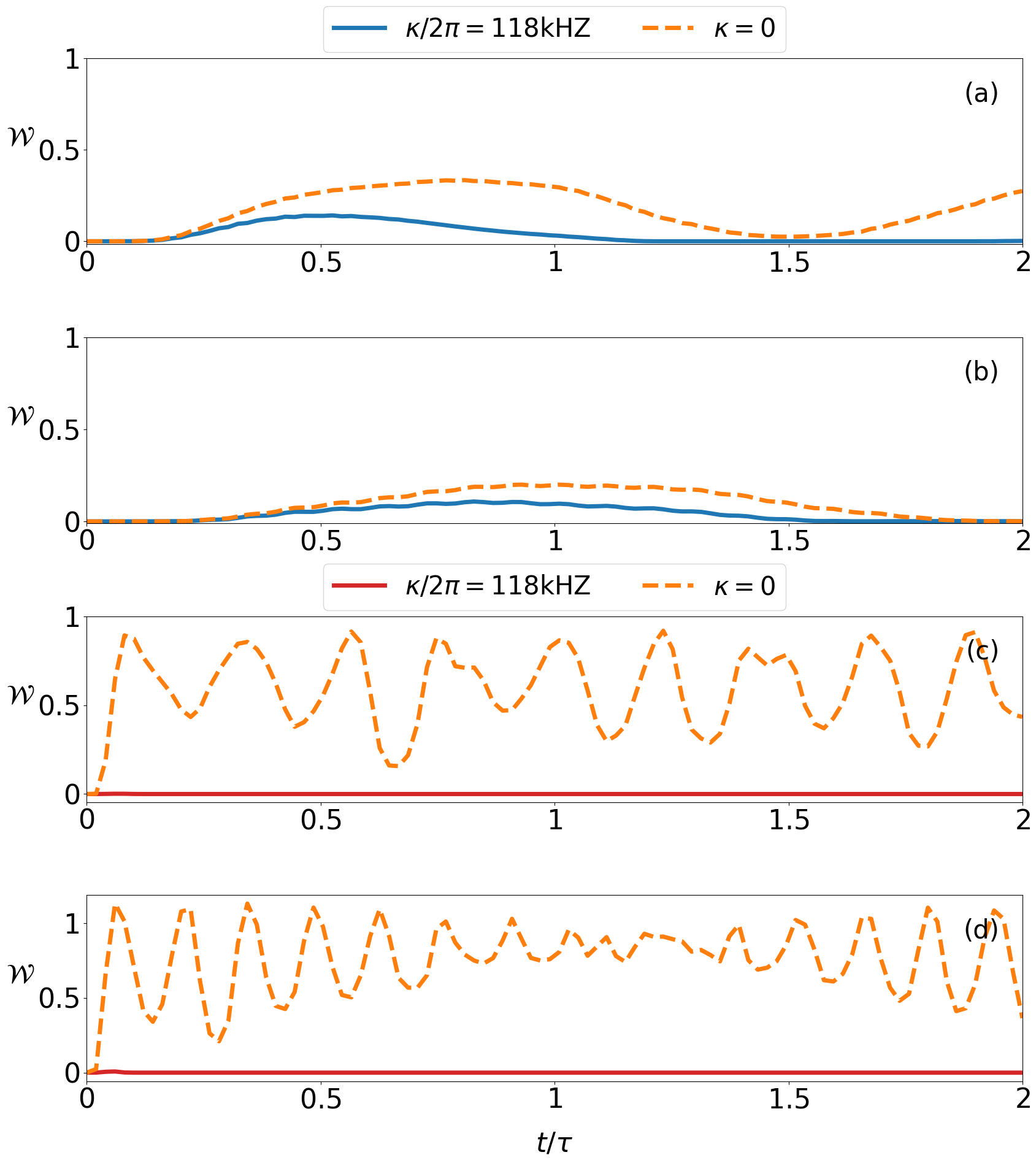}
	\caption{
	Time evolution of the Wigner function negativity for
	(a) $C_K=0.05, \widetilde C_\epsilon=0.1$,
	(b) $C_K=0.1, \widetilde C_\epsilon=0.1$,
	(c) $C_K=10^{-3}, \widetilde C_\epsilon=0.02$,
	and (d) $C_K=10^{-3}, \widetilde C_\epsilon=0.03$.
	The parameter $\tau = \pi / 2g_K$ varies across the four plots, as it depends on $C_K$, which differs in each case.
		The other system parameters are set as $\omega_m/2\pi=5.33 \text{MHz}$, $\bar{\Delta}_c=\omega_m$, and $C_E=10^{-2}$. 
	}
	\label{fig8}
\end{figure}

To investigate the influence of system dissipation on the Wigner functions of the generated nonclassical states, and to understand the degree of their nonclassicality, we use the Wigner negativity $\mathcal{W}$, which is defined as
 \cite{kenfack2004}
\begin{equation}\label{Wigner-negativity}
\mathcal{W}=\iint[\abs{W(x,p)}-W(x,p)]dxdp.
\end{equation} 
Figure~\ref{fig8} shows numerically calculated time evolution of the Wigner negativity in the presence of cavity dissipation, with $ \kappa/2\pi=118$ kHz (solid lines), and also for the ideal (free-dissipation) case $\kappa=0$ (dashed lines).

In the absence of dissipation, the \textit{Wigner negativity} attains higher values and exhibits temporal fluctuations with respect to the normalized time \( t/\tau \), indicating the sustained presence of nonclassical characteristics. In contrast, under dissipative conditions, decoherence effects gradually become dominant, eventually leading to the suppression of Wigner negativity. This behavior reflects the degradation of nonclassicality due to the interaction with the environment.

The strong-coupling regime, as illustrated in Figs.~\ref{fig8}(a) and (b), facilitates the generation of nonclassical states that display temporary robustness against dissipation, although this resilience is limited to timescales shorter than \( \tau \). On the other hand, Figs.~\ref{fig8}(c) and (d) show that nonclassical features deteriorate more rapidly in these regimes, indicating a higher sensitivity to environmental decoherence. This observation is consistent with Figs.~\ref{fig7}(c) and (d), where the chosen parameter sets fail to support the formation of well-defined nonclassical states.

Interestingly, in the dissipation-free case, Figs.~\ref{fig8}(c) and (d) exhibit oscillatory dynamics with maximum Wigner negativity values approaching unity, suggesting a persistent yet less structured form of nonclassicality. In contrast, Figs.~\ref{fig8}(a) and (b) display a more stable negativity plateau around \( t = \tau \), indicating a relatively robust quantum feature. This comparison highlights that the transition from classical to nonclassical behavior is more pronounced in the strong-coupling regime at this characteristic timescale.

It is worth emphasizing that the realization of a steady-state nonclassical state requires cooling the mechanical mode near its ground state, where \( n_{th} \simeq 0 \). The presence of thermal noise rapidly destroys quantum coherence and suppresses any nonclassical features of the system.


\section{experimental realization of Model}\label{Section7}
The experimental realization of the setup considered in Fig.~\ref{fig1}, has been reported in Ref.~\cite{Toth2017}.
The mechanical resonator contains a parallel-plate capacitor with resonance frequency $\omega_m/2\pi=5.33 \text{MHz}$ with decay rate of mechanical mode $\gamma_m/2\pi=30 \text{Hz}$, and effective decay rate $\Gamma_m/2\pi=500 \text{kHz}$. Also, the total decay rate of primary mode is $\kappa/2\pi=118 \text{kHz}$ which is the combination of internal and external decays as $\kappa^{\rm in}/2\pi=76 \text{kHz}$ and $\kappa^{\rm ex}/2\pi=42 \text{kHz}$, respectively. The laser frequency is approximately $\omega_L\approx1\text{GHz}$.

One of the key challenges in our proposal is determining the value of the optomechanical coupling. By selecting $\bar{\Delta}_c = \omega_m$, the coupling ratio is expressed as $C_K = (g_0/\omega_m)^2$, and the values of this parameter we have considered lie in the range $C_K \approx 10^{-4}-1$. The vacuum electromechanical coupling strength can be derived from the formula \cite{Toth2017, thesis-bernier}:
\begin{equation}\label{g_0}
(\frac{g_0}{\omega_m})^2=\frac{1}{4\bar n_m}(\frac{\kappa}{\kappa_{\rm ex}})^2\frac{P_{\rm cal}}{P_{\rm in}}\frac{P^{\rm meas}_{\rm SB}}{P^{\rm meas}_{\rm cal}}(1+(\frac{\kappa}{2\omega_m})^2).
\end{equation}
Here, $P_{\rm in}$ represents the pump power from the microwave source attenuated by an insertion loss or variable microwave attenuator before reaching the cavity. 
Note that by tuning and inserting attenuators, one can decrease and adjust our desired values for $ P_{\rm in} $. The sideband emission, measured as $P^{\rm meas}_{\rm SB}$, is amplified by an overall gain. To eliminate uncertainties from unknown variables, a weak calibration tone $P_{\rm cal}$ is introduced, following the same path as the pump tone. This calibration tone is tuned near the sideband at $\omega \approx \omega_m$, and the corresponding calibration power $P^{\rm meas}_{\rm cal}$ is measured.
The initial coupling strength for the given parameters, as reported in Ref.~\cite{Toth2017}, is $g_0^{\rm int} = 2\pi \times 106$ Hz.

For different parameter sets to attain the desired range of $C_K$ values in this paper, the ratio of the coupling strengths must reach $g_0/g_0^{\rm int} \approx 10^{3} - 10^{6}$. Focusing on the most relevant parameter regime for our results, we consider $g_0 / g_0^{\rm int} \sim 10^{3}\!-\!10^{4}$, which appears more accessible using current state-of-the-art circuit--optomechanical platforms. Importantly, the nonclassical features we predict already emerge in this window: the negative Mandel $Q$ and negative Wigner-function regions (Figs.~\ref{fig4} and~\ref{fig7}) occur at $C_K = 0.05$ ($g_0 / g_0^{\rm int} \approx 10^{4}$, $g_0 \approx 10^{6}\,\mathrm{Hz}$), while the squeezed states in Fig.~6 appear at $C_K = 10^{-3}$ ($g_0 / g_0^{\rm int} \approx 10^{3}$, $g_0 \approx 10^{5}\,\mathrm{Hz}$). These values indicate that the predicted nonclassical behavior is attainable with feasible experimental parameters, which naturally raises the question of how the required ratio $g_0/g_0^{\rm int}$ can be realized in practice.
Firstly, through understanding the system parameters that govern this ratio, and secondly, through practical strategies that directly increase $g_0$ in realistic circuit--optomechanical devices.

\begin{figure}
	\centering
	\includegraphics[width=\linewidth]{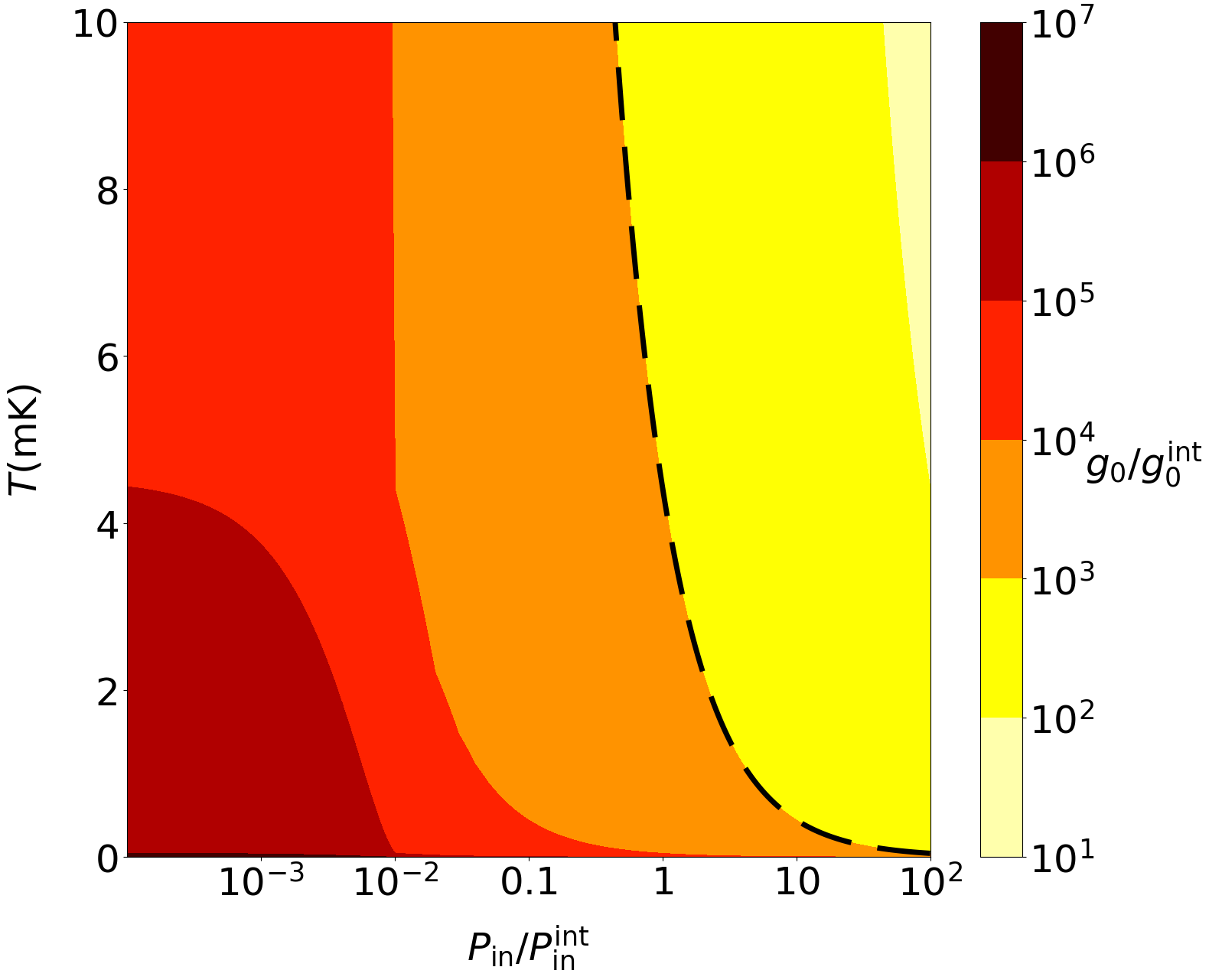}
	\caption{
		Contour plot of the ratio of vacuum optomechanical coupling versus temperature and the ratio of pump power. The black dash line correspond to $g_0/g_0^{\rm int}=10^3$. In the region left side of the black-dash curve one can achieve the required coupling strength for the desired regime.
			}
	\label{fig9}
\end{figure}

To address the first aspect, Fig.~\ref{fig9}, obtained using Eq.~\eqref{g_0}, illustrates how the ratio $g_0/g_0^{\rm int}$ depends on both the temperature and the pump power $P_{\rm in}/P_{\rm in}^{\rm int}$, where $P_{\rm in}^{\rm int}$ is defined by the intrinsic coupling $g_0^{\rm int}$. As shown, reducing either the pump power or the temperature helps to reach the required coupling strength $g_0$ for accessing the desired regime (region left of the black dashed line). The key parameter controlling $g_0/g_0^{\rm int}$ is the temperature: ultralow temperatures well below $10\,\mathrm{mK}$, already achievable in modern cryogenic systems, provide favorable conditions to obtain $g_0/g_0^{\rm int} \sim 10^{3}\!-\!10^{4}$, which is sufficient for observing several of the nonclassical effects discussed in this work \cite{Cattiaux2021}.

Secondly, as a direct way to increase $g_0$, ongoing efforts in mm-wave optomechanical circuits exploit higher cavity frequencies \cite{Miranda2025, Larson2025, Genter2024}. In addition, experimental groups are exploring high-mechanical-stress superconducting films, such as Niobium and Niobium Titanate, with stresses exceeding 1 GPa. By reducing the capacitor gap below 10 nm, these approaches can achieve up to a 100-fold increase in the intrinsic coupling rate $g_0$, providing a complementary path to reach stronger coupling regimes \cite{Youssefi2025}. Together with precise control of temperature, pump power, and device fabrication parameters, these strategies offer practical routes to access stronger coupling regimes and to observe the nonclassical phenomena predicted in this work.

The device is mounted on the base plate of a dilution refrigerator, cooled to a base temperature of approximately 10 mK, and further temperature reduction is required to enhance the optomechanical coupling. Additionally, the microwave input lines are heavily attenuated to suppress residual thermal noise, and filter cavities are employed to eliminate unwanted frequency noise from the applied tones \cite{Toth2017}.

Also, the laser amplitude is $E_L=53\text{kHz}$ which implies the values of the driving laser power have been set $ P_L\approx-190 \text{dBm}$. Moreover, we have set the weak-modulation parameter as $\epsilon\approx0.01-0.1$ which implies $\epsilon_L\approx10-100\mu\text{Hz}$.
Other parameters, although not directly used in calculations, should be specified to establish their experimental range. 
The intermode coupling strength $\frac{J}{4\pi}\vert \omega_{\rm aux}-\omega_c \vert=0.57 \text{GHz}$, and also the mechanical mode couples to both the primary and auxiliary cavity modes with an initial vacuum electromechanical coupling strength $g_0/2\pi=2\times 60 \text{Hz}$. 
Additionally, the internal and external dissipation rates are, respectively, $\kappa_{\rm aux}^{\rm ex}/2\pi=4233$ kHz, and $\kappa_{\rm aux}^{\rm in}/2\pi=245$ kHz which implies the total decay rate of auxiliary being  $\kappa_{\rm aux}/2\pi=4487 \text{kHz}$, which clearly shows that the critical condition of $\kappa\ll\kappa_{\rm aux}$ for RDR is satisfied. 
Finally, the resonance frequency of the primary and auxiliary modes are $\omega_c/2\pi =4.26 \text{GHz}$ and $\omega_{\rm aux}=5.48 \text{GHz}$, respectively.
It can be noted that there is another manufactured device with different experimental values which is reported in Ref.~\cite{Toth2017}.

After amplification with a commercial high-electron-mobility transistor (HEMT) amplifier mounted on the $ 3 ~ K $ plate, the signal can be measured with an electromagnetic spectrum analyzer (ESA) or a vector network analyzer (VNA). 
Based on the optomechanical sideband cooling by pumping the auxiliary mode corresponding
to a mean intracavity photon number $ n_{aux}\sim 1.5 \times 10^8 $ on the lower motional sideband, the mechanical oscillator can be prepared as a strongly dissipative-cold reservoir, while still remaining in the weak-coupling regime for the auxiliary mode \cite{Toth2017}.
Although this order of photons can elevate the mechanical cavity's temperature and increase thermal photons, this loss can be mitigated \cite{Yousefi2023}.
Also, a dissipative mechanical reservoir for the primary, high-Q mode is realized since $ \Gamma_m \gg \kappa $.

Finally, it should be noted that employing the fast-time modulation technique allows us to surpass the experimental challenge of frequency matching in conventional DCE. 
To conclude, we emphasize implementing our proposed scheme with current state-of-the-art technology are feasible by precisely controlling the cryostat temperature, adjusting the initial microwave pump power $P_{\rm in}$, and fabricating devices in the bad cavity (unresolved sideband, $\kappa/\omega_m >1$) regime.


\section{Conclusion Remarks}\label{Section8}

We proposed a feasible experimental optomechanical scheme in the RDR, where the mechanical damping rate greatly exceeds the cavity damping rate, in order to emerge controllable optomechanically-Kerr-assisted parametric-DCE of microwave-photons via weak coherent time modulation of the driving laser frequency. We show that our full analytical solutions under several experimentally applicable approximations are in a good agreement with the numerical results. 
The effective Hamiltonian of the system not only includes the parametric-DCE due to the coherent time modulation, but also includes optomechanically induced Kerr nonlinearity plus two extra nonlinear terms originated from the optomechanical interaction which significantly affect on the nonclassical properties of the generated Casimir photons that never occurs in the conventional DCE schemes. 
The induced nonlinear terms are responsible for the oscillatory behavior and saturation dynamics of the mean number of the generated DCE photons quadrature squeezing and sub-Poissonian photon counting statistics. 
All these properties can be controlled thorough the tunable experimental parameters such as the depth of time-modulation ($ \epsilon_L $), cryostat temperature of mechanics, and initial microwave pump assisted by variable attenuator ($ P_{\rm in} $) as well as the fabrication device parameter such as ratio $ \kappa/\omega_m $.

One of the most significant results is the generation of Casimir photons with sub-Poissonian statistics and a negative Wigner function, both of which are nonclassical features observed in a specific regime.
Notably, this system stands out as it successfully achieves both of these characteristics within the same regime.

For comparison, we note that earlier studies have typically reported either sub-Poissonian photon statistics or Wigner-function negativity, but not both simultaneously within the same physical regime. For example, sub-Poissonian statistics and quadrature squeezing were investigated in vibrating or resonantly oscillating cavities without addressing the Wigner-function negativity \cite{Dodonov1996, Dodonov1999}. 
On the other hand, it has been shown \cite{Johansson2013,Dezael2010} that the quantum states of the generated photons in the optical and superconducting analogues of the DCE exhibit Wigner-function negativity, while they obey the (classical) super-Poissonian counting statistics. This comparison highlights one the distinctive aspects of the present work in demonstrating the coexistence of Wigner-function negativity and sub-Poissonian statistics, as nonclassical signatures, within the same physical regime.

Finally, although the effective Hamiltonian studied in this work resembles that of a Kerr nonlinear resonator, its realization in the reversed-dissipation regime of an optomechanical system provides several important advantages. Unlike conventional Kerr platforms, the effective Kerr coefficient in our scheme is tunable in situ through optomechanical parameters, allowing access to regimes of strong nonclassicality—such as simultaneous negativity of the Mandel parameter and the Wigner function—that are not available in standard implementations. This tunability, together with the integrability and versatility of circuit optomechanical devices, makes our platform a promising candidate for generating nonclassical microwave radiation for quantum computing and quantum microwave sensing applications.

\section*{Acknowledgments}
AMF and HS would thank the ICQT and the CQST, respectively, for their supports. 
\textit{Contributions}. AMF proposed and developed the primary idea of DCE in RDR-optomechanics. Also, AMF performed the analytical solution in the short approximation, and then have been checked by HS. MHN provided the method for finding the analog full Hamiltonian of (\ref{HA}). Analytical calculation for finding analog Hamiltonian is performed by HS. All numerical calculations have been done by HS. Experimental discussion is developed by AMF, completed by HS and checked by AY. AY checked the suggested experimental scheme based on the current feasible technologies. RR contributed to discussion related to analog DCE Hamiltonian of the system. The initial manuscript has been written by HS and AMF, and revised by MHN. All authors contributed to prepare the manuscript and revised version based on the referees comments.
\appendix

\section{\texorpdfstring{Derivation of the Hamiltonian \eqref{H2}}{Derivation of the Hamiltonian (X)}}
\label{Appendix_A}

The Hamiltonian of the subsystem composed of the driven auxiliary mode and the MO in the frame rotating with the frequency $\omega_L^{\rm aux}$ can be written as  
\begin{equation}\label{H_aux}
\hat H'= -\Delta_{\rm aux} \hat{a}_s^\dagger \hat{a}_s + \omega_m \hat b^\dag \hat b - g_0 \hat{a}_s^\dagger \hat{a}_s (\hat b+ \hat b^\dag) + i E_L^{\rm aux} (\hat{a}_s^\dagger  - \hat{a}_s),
\end{equation}
where $\Delta_{\rm aux} = \omega_L^{\rm aux}-\omega_{\rm aux}$ stands for the detuning between the auxiliary mode and its driving field.

If the auxiliary mode is intensely driven so that the intracavity field is strong, which is realized for high-finesse cavities and enough driving power, the quantum Langevin equations can be solved analytically by adopting a linearization scheme in which the operators (here, the operators of the auxiliary mode) are expressed as the sum of their classical mean values and small fluctuations, $ \hat a_s = \bar a_s + \delta \hat a_s $ with $ \langle \delta \hat a_s^\dag \delta \hat a_s \rangle / \langle \hat a_s^\dag \hat a_s \rangle \ll 1  $. Therefore, the linearized quantum Langevin equations are given by 
\begin{eqnarray} 
&&  \delta\dot{\hat{a}}_s = (i \Delta_{\rm aux} - \frac{\kappa_{\rm aux}}{2}) \delta \hat a_s + i g' (\hat b  + \hat b^\dag) +  \sqrt{\kappa_{\rm aux}} \hat a_{\rm in}^{\rm aux}, 
\label{langevi1_appendix} 
\\
&&\dot {\hat b} =  -(i\omega_m  + \frac{\gamma_m}{2}) \hat b +  ig' ( \delta \hat a_s^\dag + \delta \hat a_s )+ \sqrt{\gamma_m} \hat b_{\rm in},
\label{langevi2_appendix} 
\end{eqnarray}
where $ g'=g_0 \bar{a}_s $ is the enhanced optomechanical coupling, and $\hat{a}^{\rm aux}_{\rm in}$ and $\hat{b}_{\rm in}$ denote, respectively, the input vacuum noise for the auxiliary mode and the Brownian noise associated with the coupling of the MO to the thermal environment, characterized by the following non-vanishing correlation functions \cite{Gardiner2000}.
\begin{equation}\label{Noise_a}
\langle {{{\hat a}_{\rm in}^{\rm aux}}(t)\hat a_{\rm in}^{\rm aux\dag} (t')}\rangle = \delta (t - t'),
\end{equation}
\begin{equation}\label{Noise_b}
\langle {{{\hat b}_{\rm in}^\dagger}(t)\hat b_{\rm in} (t')}\rangle = \bar{n}_m\delta (t - t'), \quad \langle {{{\hat b}_{\rm in}}(t)\hat b_{\rm in}^\dagger (t')}\rangle = (\bar{n}_m+1)\delta (t - t'),
\end{equation}
where $\bar{n}_m=[\exp(\hbar\omega_m/k_BT)-1]^{-1}$ is the mean thermal excitation number of the MO at temperature $T$.

The Langevin equations of motion~\eqref{langevi1_appendix} and~\eqref{langevi2_appendix} can be treated as algebraic equations by moving to Fourier space:
\begin{eqnarray}
\delta \hat{a}_s[\omega]=&&ig'\frac{\kappa_{\rm aux}/2+i(\Delta_{\rm aux}+\omega)}{\kappa_{\rm aux}^2/4+(\Delta_{\rm aux}+\omega)^2}(\hat{b}[\omega]+\hat{b}^\dagger[\omega])
\nonumber
\\
&&\sqrt{\kappa_{\rm aux}}\frac{\kappa_{\rm aux}/2+i(\Delta_{\rm aux}+\omega)}{\kappa_{\rm aux}^2/4+(\Delta_{\rm aux}+\omega)^2}\hat a_{\rm in}^{\rm aux}[\omega],
\end{eqnarray}
\begin{eqnarray}
-i\omega\hat{b}[\omega]=&&-(i\omega_m  + \frac{\gamma_m}{2}) \hat b[\omega] +  ig' ( \delta \hat a_s^\dag[\omega] + \delta \hat a_s[\omega] )
\nonumber
\\
&&+ \sqrt{\gamma_m} \hat b_{\rm in}[\omega].
\end{eqnarray}
Combining these two equations leads to
\begin{eqnarray}\label{b_combination}
\hat b[\omega] =&&\big(i\frac{\Gamma_m}{2}-\frac{\Omega_m}{\omega}
\big)
\hat b[\omega] + \big(i\frac{\Gamma_{\rm op}}{2}-
\frac{\omega_{\rm op}}{\omega}
\big)\hat b^\dagger[\omega]
+ \sqrt{\gamma_m}\hat{b}_{\rm in}[\omega]
\nonumber
\\
&&
+
\sqrt{\kappa_{\rm aux}}\Big[
\frac{\kappa_{\rm aux}/2+i(\Delta_{\rm aux}+\omega)}{\kappa_{\rm aux}^2/4+(\Delta_{\rm aux}+\omega)^2}\hat{a}_{\rm in}^{\rm aux}[\omega]
\nonumber
\\
&&-\frac{\kappa_{\rm aux}/2+i(\Delta_{\rm aux}-\omega)}{\kappa_{\rm aux}^2/4+(\Delta_{\rm aux}-\omega)^2}
\hat{a}_{\rm in}^{\rm aux \dagger}[\omega]
\Big],
\end{eqnarray}
where 
\begin{eqnarray} \label{Gamma-Omega}
&& \Omega_m \equiv \omega_m + \omega_{\rm op} , \\
&& \Gamma_m \equiv \gamma_m + \Gamma_{\rm op},
\end{eqnarray}
with
\begin{eqnarray}
\omega_{\rm op}[\omega] = g'^2 \left[ \frac{\Delta_{\rm aux} + \omega}{\kappa_{\rm aux}^2/4 + (\Delta_{\rm aux}+ \omega)^2} + \frac{\Delta_{\rm aux} - \omega}{\kappa_{\rm aux}^2/4 + (\Delta_{\rm aux}   - \omega)^2 } \right],
\nonumber \\ \label{omega-op} 
\end{eqnarray}
\begin{eqnarray}
\Gamma_{\rm op}[\omega] = g'^2  \left[ \frac{\kappa_{\rm aux}}{\kappa_{\rm aux}^2/4 + (\Delta_{\rm aux} + \omega)^2} - \frac{\kappa_{\rm aux}}{\kappa_{\rm aux}^2/4 + (\Delta_{\rm aux} - \omega)^2 } \right],\nonumber \\
\end{eqnarray}
are, respectively, referred to as the shifted mechanical frequency and effective mechanical damping rate due to the optomechanical interaction \cite{Aspelmeyer2014}. Hence, the auxiliary mode with higher damping rate renormalizes both the frequency and the damping rate of the MO.

In the red-detuned regime $\Delta_{\rm aux}=-\omega_m$, we have
\begin{eqnarray} \label{omega-gamma_cooling}
&& \Gamma_{\rm op}[\omega_m]= 4g'^2 \left(\frac{1}{\kappa_{\rm aux}} - \frac{\kappa_{\rm aux} }{\kappa_{\rm aux}^2 + 16 \omega_m^{2}} \right), 
\\
&& \omega_{\rm op}[\omega_m] = - \frac{2g'^2\omega_m}{\kappa_{\rm aux}^2/4 + 4 \omega_m^{2}}. 
\end{eqnarray}
The maximum amount of optomechanical cooling can be reached in the\textit{ resolved-sideband }regime where $ \omega_m~\gg~\kappa_{\rm aux}$. In this regime the induced optomechanical frequency and induced optomechanical damping rate are respectively,  $\omega_{\rm op} / \omega_m \simeq -0.5 (g'/\omega_m)^2$ and $\Gamma_{\rm op}/\kappa_{\rm aux} \simeq 4 (g'/\kappa_{\rm aux})^2 $. Therefore, the shifted mechanical frequency and the effective damping rate in the resolved-sideband regime are, respectively, given by
\begin{eqnarray} 
&& \Omega_m = \omega_m (1-\frac{1}{2} \frac{g'^2}{\omega_m^2}),  
\label{omega_resolved} 
\\
&& \Gamma_m=\gamma_m (1+\mathcal{C}), 
\label{gamma_resolved}
\end{eqnarray}
where $ \mathcal{C}=4g'^2/\kappa_{\rm aux}\gamma_m $ is the optomechanical cooperativity. It is clear that in the strong coupling regime $ g' \gg \kappa_{\rm aux} $ or large cooperativity $ \mathcal{C} \gg 1 $, the effective mechanical damping rate can be adjusted to be very large by controlling the optomechanical cooperativity, i.e., $ \Gamma_m \gg \kappa,\kappa_{\rm aux} $. Also, in the parameter regime where $ g' \ll \omega_m $, the shifted mechanical frequency is approximately equal to natural mechanical frequency $ \Omega_m \approx \omega_m $.

Since the damping rate of the auxiliary mode is much greater than that of the MO $(\kappa_{\rm aux}\gg \gamma_m)$ one can adiabatically eliminate the auxiliary mode $\hat{a}_s$ on time scales longer than $\kappa_{\rm aux}^{-1}$. Using $\Gamma_m\gg\Gamma_{\rm op}$ and $\omega_m\gg\omega_{\rm op}$ in Eq.~\eqref{b_combination}, and then Fourier transforming back into the time domain we obtain
\begin{equation}\label{b_final_appendix}
\dot {\hat b} \approx -i \Omega_m  \hat b - \frac{\Gamma_m}{2} \hat b + \sqrt{\Gamma_m} \hat {\tilde b}_{\rm in}, 
\end{equation}
where
\begin{eqnarray}
\sqrt{\Gamma_m}  \hat{\tilde{b}}_{\rm in}= \sqrt{\gamma_m} \hat b_{\rm in} + 
\frac{\kappa_{\rm aux}/2+i2\Delta_{\rm aux}}{\kappa_{\rm aux}^2/4+4\Delta_{\rm aux}^2}
\hat{a}_{\rm in}^{\rm aux}
-\frac{2}{\sqrt{\kappa_{\rm aux}}} 
\hat{a}_{\rm in}^{\rm aux \dagger}.
\nonumber
\\ 
\label{b_in}
\end{eqnarray}
is identified as a generalized mechanical noise operator. In this way, from Eq.~\eqref{b_final_appendix}, one can easily deduce the Hamiltonian of Eq.~\eqref{H2} in the frame rotating with the driving laser frequency $\omega_L$.


\section{Time evolution operator in the WCR}\label{Appendix_B}
In order to derive an analytical expression for the system time-evolution operator in the WCR and in the absence of system dissipation, we first transform the Hamiltonian~\eqref{H_WR} to a frame defined by the unitary transformation $U_0={\rm exp} (-i\bar\Delta_c t \hat n)$,
\begin{gather} \label{H_eff-mod1}
\hat H_{\rm WCR}' =U_0^\dagger\hat H_{\rm WCR}U_0=
\nonumber
\\
\bar\Delta_c\big[ -C_K \hat a^{\dag 2} \hat a^2 + i C_\epsilon(t)(\hat a^{\dag 2}e^{i2 \bar \Delta_c t } - \hat a^2e^{-i2 \bar \Delta_c t })
\nonumber
\\
+ i C_E (\hat a^\dag e^{i \bar \Delta_c t } -\hat a e^{-i\bar \Delta_c t})\big].
\end{gather}
Using Eq.~\eqref{C_E}, we obtain
\begin{gather} \label{H_eff-mod2}
\hat H_{\rm WCR}'/\bar{\Delta}_c = -C_K \hat a^{\dag 2} \hat a^2 
+ i \frac{\widetilde{C_\epsilon}}{2}(\hat a^{\dag 2}-\hat a^2)
\nonumber
\\
+ i C_E (\hat a^\dag e^{i \bar \Delta_c t } -\hat a e^{-i\bar \Delta_c t})
+ i\frac{\widetilde{C_\epsilon}}{2}(\hat a^{\dag 2}e^{i4 \bar \Delta_c t } - \hat a^2e^{-i4 \bar \Delta_c t }).
\end{gather}

By using the rotating wave approximation to drop the terms proportional to $e^{\pm i\bar \Delta_c t}$ and $e^{\pm i4\bar \Delta_c t}$ which is valid for $\bar \Delta_c^{-1}\ll t$, the total Hamiltonian in the rotating frame is given by 
\begin{eqnarray} \label{H_eff-mod3}
&& \hat H_{\rm WCR}'= -g_K  \hat a^{\dag 2} \hat a^2 +i\chi'(\hat a^{\dag 2} - \hat a^2),
\end{eqnarray}
where $\chi'=\frac{\widetilde{C}_\epsilon}{2}\bar{\Delta}_c$.
We proceed by transforming the Hamiltonian \eqref{H_eff-mod3} into the interaction picture generated by the unitary operator $ \hat U_1=e^{-ig_K \hat n^2 t} $. Thus, we get
\begin{eqnarray} \label{H_eff-mod4}
\mathcal{\hat H}_{\rm eff}=&&\hat U_1^\dag \hat H_{\rm WCR}'\hat U_1=-g_K \hat n 
\nonumber
\\
&&+ i \chi' \left( \hat a^{\dag 2} e^{4ig_Kt(\hat n+1)}  - e^{-4ig_K t(\hat n+1)} \hat a^2 \right).
\end{eqnarray}
Introducing three operators $ \hat L_{\pm,z} $ and the function $ f(t) $ as follows
\begin{subequations}
	\begin{eqnarray}
	&& \hat L_+(t)=\frac{1}{2} \hat a^{\dag 2} e^{4ig_K \hat n t}=\hat L_-^\dagger(t), \label{L+-}\\
	&&  \hat L_z= \frac{1}{2} (\hat n+ \frac{1}{2}), \label{Lz} \\
	&& f(t)=2i\chi' e^{i4g_K t},
	\end{eqnarray}
\end{subequations}
we can rewrite the explicit time-dependent Hamiltonian~\eqref{H_eff-mod4} as follows
\begin{equation} \label{H_eff-mod5}
\mathcal{\hat H}_{\rm eff}(t)= -2g_K \hat L_z + f(t) \hat L_+(t) + f(t)^\ast \hat L_-(t)+\frac{g_K}{2} ,
\end{equation}
where the last term does not influence the dynamics of the system except for a global phase factor $ e^{-ig_Kt/2} $. The first three terms satisfy the \textit{su}(1,1) Lie algebra with commutation relations $ [\hat L_-,\hat L_+]=2\hat L_z $ and $ [\hat L_z,\hat L_{\pm}]=\pm \hat L_{\pm} $. For an infinitesimally short interval of time $ \delta t $, the time evolution operator is
\begin{eqnarray}\label{U}
 \hat U_{I} \vert_{\delta t \to 0}  \approx e^{-i\mathcal{\hat H}_{\rm eff}  \delta t} = {\rm exp}[a_z \hat L_z + a_+ \hat L_+(t) + a_- \hat L_-(t)],
\end{eqnarray} 
where $ a_z=2i g_K \delta t $, $ a_+=-i \delta t f(t) $ and $ a_-=-i \delta t f(t)^\ast $.
Using the normal-order decomposition formula for exponent functions of the generators of the \textit{su}(1,1) Lie algebra we arrived at
\begin{gather}\label{U_I}
\hat U_I = e^{-i\mathcal{\hat H}_{\rm eff}\delta t}  = e^{A_+(t) \hat L_+(t)} e^{\ln(A_z(t)) \hat L_z} e^{A_-(t) \hat L_-(t)},
\end{gather}
where the complex time-dependent functions $A_\pm$ and $A_z$ are given by
\begin{eqnarray} \label{normal function}
&& A_{\pm }(t)= \frac{(a_{\pm}/\phi)\sinh \phi}{\cosh \phi - (a_z/2 \phi) \sinh \phi}, \\
&& A_z(t)=[\cosh \phi - (a_z/2\phi)\sinh \phi]^{-2},
\end{eqnarray}
with $\phi=\sqrt{(a_z/2)^2 - a_+ a_-}$. 
For a finite-time interval $0$ to $t$, the time evolution operator can be written as
\begin{eqnarray} \label{UI}
\hat U_I(t)= \underset{\delta t \to 0}{\mathop{\lim }}\,\mathcal{\hat T}\prod\limits_{j=0}^{t/\delta t} e^{A_+(t_j) \hat L_+(t_j)} e^{A_z(t_j) \hat L_z} e^{A_-(t_j) \hat L_-(t_j)} ,
\end{eqnarray}
where $ t_j=j \delta t $ and $ \mathcal{\hat T} $ is the time ordering operator. 
The time dependence of the generators $\hat{L}_\pm(t)$ through the exponential terms $e^{\pm 4ig_K\hat{n}t}$ makes complicate to find an efficient way to apply the time evolution operator $\hat{U}_I(t) $ on an arbitrary initial state. To circumvent this difficulty, we take \cite{Roman2015} $e^{\pm4ig_K\expval{\hat{n}}t}\!\approx1$, which is justified for times such that $t\ll(4g_K\expval{n})^{-1}$.
Under this approximation, the generators $\hat{L}_\pm$ appeared in the effective Hamiltonian~\eqref{H_eff-mod5} become time-independent. Now, by applying the Wei-Norman theorem \cite{Norman1964} we can express the time evolution operator in the form of a product of exponentials as follows
\begin{equation}\label{u_norman}
\mathcal{\hat U}(t) \approx e^{\alpha(t) \hat L_+} e^{\beta (t) \hat L_z} e^{\gamma (t) \hat L_-}.
\end{equation}
The time-dependent functions $\alpha(t),\beta(t),$ and $\gamma(t)$ can be determined by substituting the time evolution operator~\eqref{u_norman} into the time-dependent Schrödinger equation in the interaction representation. In doing so, we obtain a set of coupled nonlinear ordinary differential equations,
\begin{subequations} \label{dif-eq}
	\begin{equation}
	\dot \alpha(t)=-i [f(t) - 2g_K \alpha (t) + f(t)^\ast \alpha^2(t)],
	\end{equation}
	\begin{equation}
	\dot \beta (t)= -i[-2g_K+2 f(t)^\ast \alpha (t)],
	\end{equation}
	\begin{equation}
	\dot \gamma (t)= -i f(t)^\ast e^{\beta (t)},
	\end{equation}
\end{subequations}
with the initial conditions $\alpha(0)=\beta(0)=\gamma(0)=0$. 
The first equation is the well-known Ricatti differential equation which has an analytical solution as 
\begin{subequations}\label{dif-ans_app}
\begin{equation}\label{alpha_app}
\alpha(t)=\frac{2\chi' e^{4ig_K t } \sinh(\mathcal{G} t) }{\mathcal{G} \cosh(\mathcal{G} t)  + i g_K \sinh(\mathcal{G}t)}, 
\end{equation}
and two other equations can be solved by direct integration to yield
\begin{equation} \label{beta_app}
\beta(t)= 4ig_K t + 2\ln \mathcal{G} - 2 \ln [\mathcal{G} \cosh(\mathcal{G}t) + ig_K \sinh (\mathcal{G}t)],
\end{equation}
\begin{equation}\label{gamma_app}
\gamma(t)=\frac{-2 \chi'\sinh(\mathcal{G} t)}{\mathcal{G}\cosh (\mathcal{G} t) + i g_K\sinh(\mathcal{G}t) },
\end{equation}
\end{subequations}
where $ \mathcal{G}:= \sqrt{ 4\chi'^2 -g_K^2 }$. 
The total evolution operator of the system in the short-time approximation is given by 
\begin{eqnarray}\label{u_final_app}
\mathcal{\hat U}(t)= && e^{-ig_K t/2} \hat U_0 \hat U_1 \hat U_I \nonumber \\
= && {\rm exp} \left[ \frac{\beta(t)}{4}-i\frac{g_K}{2} t\right] {\rm exp}\left[-i\bar{\Delta}_c t \hat n - ig_K t \hat n^2\right]  \nonumber \\
&& {\rm exp} \left[ \frac{\alpha(t)}{2} \hat a^{\dag 2} \right] {\rm exp} \left[ \frac{\beta(t)}{2} \hat n \right] {\rm exp} \left[ \frac{\gamma(t)}{2} \hat a^{2} \right].
\end{eqnarray}
\section{\texorpdfstring{Derivation of Eq.~\eqref{n_Casimir}}{Derivation of Eq. (X)}}\label{Appendix-C}

To derive the analytic expression~\eqref{n_Casimir} for the mean number of generated Casimir photons in the WCR and in the absence of cavity dissipation, we need to determine the time-dependence of the operators $\hat{a}(t)$ and $\hat{a}^\dagger(t)$. 
For this purpose, we consider the generalized non-unitary squeezing operators $\hat{S}(\xi,\eta,\zeta)$ defined by \cite{Wunsche2003}
\begin{equation}\label{S}
\hat{S}(\xi,\eta,\zeta)=\exp[\xi\frac{1}{2}\hat{a}^2+i\eta\frac{1}{2}(\hat{a}^\dagger\hat{a}+\hat{a}\hat{a}^\dagger)-\zeta\frac{1}{2}\hat{a}^{\dagger 2}],
\end{equation}
in which $\xi$, $\eta$, and $\zeta$ are in general complex parameters. The operators $\hat{S}(\xi,\eta,\zeta)$ possess the property
\begin{equation}\label{S-properties}
[\hat{S}(\xi,\eta,\zeta)]^\dagger=\hat{S}(-\xi^*,-\eta^*,-\zeta^*)=[\hat{S}(\xi^*,\eta^*,\zeta^*)]^{-1}.
\end{equation}
The totality of these operators forms a Lie group that is the complex extension of the two-dimensional unimodular group or, equivalently, of the two-dimensional symplectic group (SL(2,C)$\sim$Sp(2,C)). 
The fundamental two-dimensional representation of this group in the basis of the annihilation and creation boson operators $\hat{a}$ and $\hat{a}^\dagger$ is obtained by the following similarity transformations of these operators \cite{Wunsche2003}
\begin{subequations}\label{Sa}
\begin{gather}\label{Sa1}
\hat{S}(\xi,\eta,\zeta)(\hat{a},\hat{a}^\dagger)[\hat{S}(\xi,\eta,\zeta)]^{-1}=(\hat{a},\hat{a}^\dagger)
\begin{pmatrix}
\kappa & \lambda
\\
\mu & \nu
\end{pmatrix}
\nonumber
\\
=(\kappa\hat{a}+\mu\hat{a}^\dagger,\lambda\hat{a}+\nu\hat{a}^\dagger).
\end{gather}
\begin{gather}\label{Sa2}
[\hat{S}(\xi,\eta,\zeta)]^{-1}(\hat{a},\hat{a}^\dagger)\hat{S}(\xi,\eta,\zeta)=(\hat{a},\hat{a}^\dagger)
\begin{pmatrix}
\nu & -\lambda
\\
-\mu & \kappa
\end{pmatrix}
\nonumber
\\
=(\nu\hat{a}-\mu\hat{a}^\dagger,-\lambda\hat{a}+\kappa\hat{a}^\dagger).
\end{gather}
\end{subequations}
The unimodular matrix of this representation is given by
\begin{equation}\label{matrix}
\begin{pmatrix}
\kappa & \lambda
\\
\mu & \nu
\end{pmatrix}
=
\begin{pmatrix}
\cosh(\mathcal{E})-i\eta r_\mathcal{E} & \xi r_\mathcal{E}
\\
\zeta r_\mathcal{E}  & \cosh(\mathcal{E})+i\eta r_\mathcal{E}
\end{pmatrix},
\end{equation}
where $r_\mathcal{E}=\frac{\sinh(\mathcal{E})}{\mathcal{E}}$  with $\mathcal{E}=\sqrt{\xi\zeta-\eta^2}$. In our case, by using Eqs.~\eqref{U},~\eqref{dif-ans_app}, and~\eqref{matrix}, we find that the connections between the two sets of parameters $(\kappa(t),\mu(t),\lambda(t),\nu(t))$ and $(\alpha(t),\beta(t),\gamma(t))$ are given by the following relations
\begin{subequations}\label{Coefficient}
	\begin{equation}
	\kappa(t)= e^{-\beta(t)/2}, \label{kappa_t}
	\end{equation}
	\begin{equation}
	\mu(t)=-\alpha(t) e^{-\beta(t)/2}, \label{mu_t} 
	\end{equation}
	\begin{equation}
	\lambda(t)= \gamma(t) e^{-\beta(t)/2} ,\label{lambda_t}
	\end{equation}
	\begin{equation}
	\nu(t)= \frac{1-\alpha(t) \gamma(t) e^{-\beta(t)}}{e^{-\beta(t)/2}}. \label{nu_t}
	\end{equation}
\end{subequations} 
In this way, the number operator $\hat{n}(t)=\hat{a}^\dagger(t)\hat{a}(t)$ is obtained as
\begin{align}\label{n1}
\hat n(t)=& \mathcal{U}^\dag \hat n \mathcal{U}=\mathcal{U}^\dag \hat a^\dag \hat a \mathcal{U} 
\nonumber \\
=& \Phi_1(t) \hat a^{\dag 2} + \Phi_2(t) \hat n + \Phi_3(t)\hat a^2 + \Phi_4(t),
\end{align}
where functions $ \Phi_i $ are given by (see Eqs.~\eqref{Sa})
\begin{subequations}\label{Phi}
	\begin{eqnarray}
	&& \Phi_1(t)= \alpha(t) e^{-\beta(t)}, \label{phi1} \\
	&& \Phi_2(t)= 1-2 \alpha(t) \gamma(t) e^{-\beta(t)}=\Phi_2^\ast(t), \label{phi2} \\
	&& \Phi_3(t)= \alpha(t) \gamma^2(t) e^{-\beta(t)} -  \gamma(t)=\Phi_1^\ast(t) , \label{phi3} \\
	&& \Phi_4(t)= -\alpha(t) \gamma(t) e^{-\beta(t)}=\Phi_4^\ast(t).
	\end{eqnarray}
\end{subequations}
It is now a simple matter to calculate the mean number of generated Casimir photons
\begin{equation}\label{n_casimir_app}
n(t)= \langle 0 \vert  \hat n(t) \vert 0 \rangle = \Phi_4(t)= \frac{4\chi'^2}{\mathcal{G}^2} \sinh^2 \mathcal{G}t.
\end{equation}

\section{\texorpdfstring{Derivation of Eqs.~\eqref{var-q} and~\eqref{var-p}}{Derivation of Eqs. (X) and (Y)}}\label{Appendix_Squeezing}
The time-evolved operators $\hat{q}^2(t)$ and $\hat{p}^2(t)$ are calculated as
\begin{eqnarray}
&& \hat q^2(t)= \mathcal{\hat U}^\dag (t) \hat q^2(0)  \mathcal{\hat U} (t)=e^{ig_K t \hat n^2(t)} \left( \hat S^{-1} \hat q^2 \hat S \right) e^{-ig_K t \hat n^2(t)}, \label{q2} \\ 
&& \hat p^2(t)=\mathcal{\hat U}^\dag (t) \hat p^2(0)  \mathcal{\hat U} (t)=e^{ig_K t \hat n^2(t)} \left( \hat S^{-1} \hat p^2 \hat S \right) e^{-ig_K t \hat n^2(t)}.  
\nonumber
\\
\label{p2}
\end{eqnarray}
By using the similarity transformation of Eq.~\eqref{Sa2} and the definition of the quadrature operators $\hat{q}$ and $\hat{p}$ we get
\begin{eqnarray}
\hat{S}^{-1}\hat{q}^2\hat{S} =&&\frac{1}{4}\Big[ \hat a^2 [\nu(t)^2 + \lambda(t)^2 + 2\Phi_3(t)]
\nonumber  
\\
&&  + \hat a^{\dag 2} [\mu(t)^2 +\kappa(t)^2 +2\Phi_1(t)]
\nonumber 
\\
&&+ 2\hat n[ \Phi_2(t) - (\mu(t) \nu(t) + \lambda(t) \kappa(t))]
\nonumber 
\\
&& +  1 + 2 \Phi_4(t) -(\mu(t) \nu(t) + \lambda(t) \kappa(t))
\Big], 
\end{eqnarray}
\begin{eqnarray}
\hat{S}^{-1}\hat{p}^2\hat{S} =&&\frac{1}{4}\Big[ -\hat a^2 [ \nu(t)^2 + \lambda(t)^2 - 2\Phi_3(t)] 
\nonumber 
\\
&&  - \hat a^{\dag 2} [ \mu(t)^2 +\kappa(t)^2 - 2\Phi_1(t)]  
\nonumber 
\\
&&  + 2\hat n [ \Phi_2(t) + ( \mu(t) \nu(t) + \lambda(t) \kappa(t))] 
\nonumber 
\\
&& 1 + 2 \Phi_4(t) + ( \mu(t) \nu(t) + \lambda(t) \kappa(t))
\Big]. 
\end{eqnarray}
Taking the vacuum state as the initial state of the cavity field, one obtains
\begin{align}\label{q2t}
\langle \hat q^2(t) \rangle &= \langle 0 \vert \hat q^2(t) \vert 0 \rangle
=
\bra{0} \hat{S}^{-1}\hat{q}^2\hat{S} \ket{0}
\nonumber
\\
&= \frac{1 +  2 \Phi_4(t)-(\mu(t) \nu(t) + \lambda(t) \kappa(t))}{4},
\end{align}
\begin{align}\label{p2t}
\langle \hat p^2(t) \rangle &= \! \langle 0 \vert \hat p^2(t) \vert 0 \rangle
=
\bra{0} \hat{S}^{-1}\hat{p}^2\hat{S} \ket{0}
\nonumber
\\
&= \frac{1 \! +  \! 2 \Phi_4(t) + \! (\mu(t) \nu(t) + \lambda(t) \kappa(t))}{4},
\end{align}
where $ \Phi_4(t)= \langle \hat n(t) \rangle $. In the short-time limit, when $ t \ll (g_K \langle \hat n \rangle)^{-1} $, we have $ \langle \hat q(t) \rangle =\langle \hat p(t) \rangle =0 $. Therefore the variances of the quadrature operators of the cavity mode are given by Eqs.~\eqref{q2t} and~\eqref{p2t}.


\end{document}